\documentclass[twocolumn,final, a4paper, 10pt]{IEEEtran}
\IEEEoverridecommandlockouts

\usepackage{latexsym}
\usepackage{graphicx}
\usepackage{comment}
\usepackage{amsfonts,amssymb,amsmath,bm}
\usepackage{algorithm}
\usepackage{relsize,exscale}
\usepackage[noend]{algorithmic}
\usepackage[dvipsnames]{xcolor}
\def\BibTeX{{\rm B\kern-.05em{\sc i\kern-.025em b}\kern-.08em T\kern-.1667em\lower.7ex\hbox{E}\kern-.125emX}}
\newcommand{\EE}{\mathbb{E}} 
\newcommand{\PP}{\mathbb{P}} 

\newcommand{\ee}{{\rm e}}
\newcommand{\jj}{{\rm j}}  
\newcommand{\dd}{{\rm\,d}} 

\newcommand{\av}{{\bf a}}

\newcommand{\ev}{{\bf e}}

\newcommand{\gv}{{\bf g}}

\newcommand{\pv}{{\bf p}}

\newcommand{\tv}{{\bf t}}
\newcommand{\uv}{{\bf u}}

\newcommand{\vv}{{\bf v}}
\newcommand{\xv}{{\bf x}}
\newcommand{\yv}{{\bf y}}

\newcommand{\zerov}{{\bf 0}}

\newcommand{\Am}{{\bf A}}

\newcommand{\Dm}{{\bf D}}

\newcommand{\Fm}{{\bf F}}

\newcommand{\Id}{{\bf I}}

\newcommand{\Um}{{\bf U}}

\newcommand{\Vm}{{\bf V}}

\newcommand{\Ac}{{\cal A}}

\newcommand{\Dc}{{\cal D}}
\newcommand{\Ec}{{\cal E}}

\newcommand{\Mc}{{\cal M}}
\newcommand{\Nc}{{\cal N}}

\newcommand{\Qc}{{\cal Q}}

\newcommand{\Tc}{{\cal T}}
\newcommand{\Uc}{{\cal U}}

\newcommand{\Vc}{{\cal V}}

\newcommand{\zetav}{\boldsymbol{\zeta}}
\newcommand{\etav}{\boldsymbol{\eta}}

\newcommand{\muv}{\boldsymbol{\mu}}
\newcommand{\mul}{\mathlarger{\mu}}
\newcommand{\muvl}{\mathlarger{\muv}}
\newcommand{\muL}{\mathlarger{\mathlarger{\mu}}}
\newcommand{\muvL}{\mathlarger{\mathlarger{\muv}}}

\newcommand{\omegav}{\boldsymbol{\omega}}

\newcommand{\Deltam}{\boldsymbol{\Delta}}

\newcommand{\Pim}{\boldsymbol{\Pi}}

\newcommand{\Omegam}{\boldsymbol{\Omega}}

\def\Tran{^{\mathsf{T}}}

\newcommand{\Dcb}{\boldsymbol{\Dc}}
\newcommand{\Vcb}{\boldsymbol{\Vc}}

\def\ben{\begin{enumerate}}
\def\beq{\begin{equation}}
\def\beqa{\begin{eqnarray}}
\def\bit{\begin{itemize}}
\def\een{\end{enumerate}}
\def\eeq{\end{equation}}
\def\eeqa{\end{eqnarray}}
\def\eit{\end{itemize}}

\def\non{\nonumber\\}

\newtheorem{remark}{Remark}

\newcommand{\aletext}[1]{{\color{black}#1}}
\newcommand{\aletextz}[1]{{\color{black}#1}}

\newcommand{\ematext}[1]{{\color{black}#1}}
\newcommand{\ematextz}[1]{{\color{black}#1}}

\newcommand{\pvk}{\bar{\pv}}
\newcommand{\vvk}{\bar{\vv}}
\newcommand{\Umk}{\bar{\Um}}

\begin{document}
\title{ Joint Communication and Sensing\\ in OTFS-based UAV Networks }

\author{Alessandro Nordio,~\IEEEmembership{Member,~IEEE,} Carla Fabiana Chiasserini,~\IEEEmembership{Fellow,~IEEE,} Emanuele Viterbo,~\IEEEmembership{Fellow,~IEEE}
\thanks{Alessandro Nordio is with CNR-IEIIT, Torino, Italy; Carla Fabiana Chiasserini is with Politecnico di Torino, Torino, Italy; Emanuele Viterbo is with Monash University, Melbourne, Australia.}
\thanks{This work was partially supported by the European Union under the Italian National Recovery and Resilience Plan (NRRP) of NextGenerationEU, partnership on “Telecommunications of the Future” (PE00000001 - program “RESTART”), and by the Project: “SoBigData.it - Strengthening the Italian RI for Social Mining and Big Data Analytics”.
E. Viterbo's work is supported by the Australian Research Council (ARC) through the Discovery project: DP200100096.}
\thanks{A preliminary version of this work was presented at Globecom'23~\cite{NoiGlobecom2023}.}
}




\maketitle
\begin{abstract}
  We consider the problem of accurately localizing $N$ unmanned aerial
  vehicles (UAV) in 3D space where the UAVs are part of a swarm and
  communicate with each other through orthogonal time-frequency space
  (OTFS) modulated signals. 
  \ematext{The OTFS communication system operates in the delay-Doppler domain and can simultaneously provide range and velocity information about the scatterers in the channels at no additional cost. }
  Each receiving UAV estimates the multipath
  wireless channel on each link formed by the line-of-sight (LoS)
  transmission and by the single reflections from the remaining $N{-}2$
  UAVs.  
  \ematextz{The estimated channel delay profiles are communicated to an
  edge server to estimate the location and velocity of the UAVs from the relative echo delay (RED) measurements between the LoS and the
  non-LoS paths. 
  To accurately obtain such estimations, we  propose a solution called Turbo Iterative Positioning (TIP), initialized by a belief-propagation approach.  Enabling a full cold start (no prior knowledge of initial positions), the belief propagation first 
  provides a map associating each echo to a reflecting UAV. The localization of the $N$ UAVs is then derived by iteratively alternating a gradient descent optimization and a refinement of the association maps between UAVs  and echos.
  Given that the OTFS receivers also acquire the Doppler shifts of each path, the UAV's velocities can be sensed jointly with communication.}
 Our numerical results, obtained also using
  real-world traces, show how the multipath links are beneficial to
  achieving very accurate position and velocity for  all UAVs, even with a
  limited delay-Doppler resolution. The robustness of our scheme is proven by its
  performance approaching the Cram\'er-Rao bound.

\end{abstract}

    











 

\section{Introduction}
Accurate localization stands as a crucial element in the repertoire
of applications that forthcoming 6G  communication networks
are anticipated to empower~\cite{Witrisal2016}. This encompasses
applications such as ensuring safety in vehicular networks and
facilitating  area exploration, rescue operations,
and relief efforts, all of which can be achieved through swarms of
unmanned aerial vehicles (UAVs)~\cite{Vehicles-survey,UAV-survey}. In
situations where GPS signals are restricted or absent, the process of
localizing these communicating nodes typically involves measuring
their distances, which is performed by estimating the time delay of
pilot signals' propagation. 

\ematext{In such demanding scenarios, it is important to integrate the communication system with a multi-target localization system in the most efficient way to provide savings in bandwidth, power, and hardware resources. This can be achieved by sharing a single waveform for both sensing and communications by leveraging the pilot-based channel estimation of the communication system to acquire the range and velocity information. 

The integration of communication and sensing  in a network environment can offer further gains but also presents additional challenges when considering synchronization requirements across all nodes. Specifically, to effectively combine the range and velocity information collected from  the nodes in the network, it is typically assumed that they all have a very accurate timing reference, down to carrier and/or symbol synchronization level. However, this is not realistic in most plesiochronous communication networks.
}

\begin{figure}[t]
  \centering
\includegraphics[width=0.95\columnwidth]{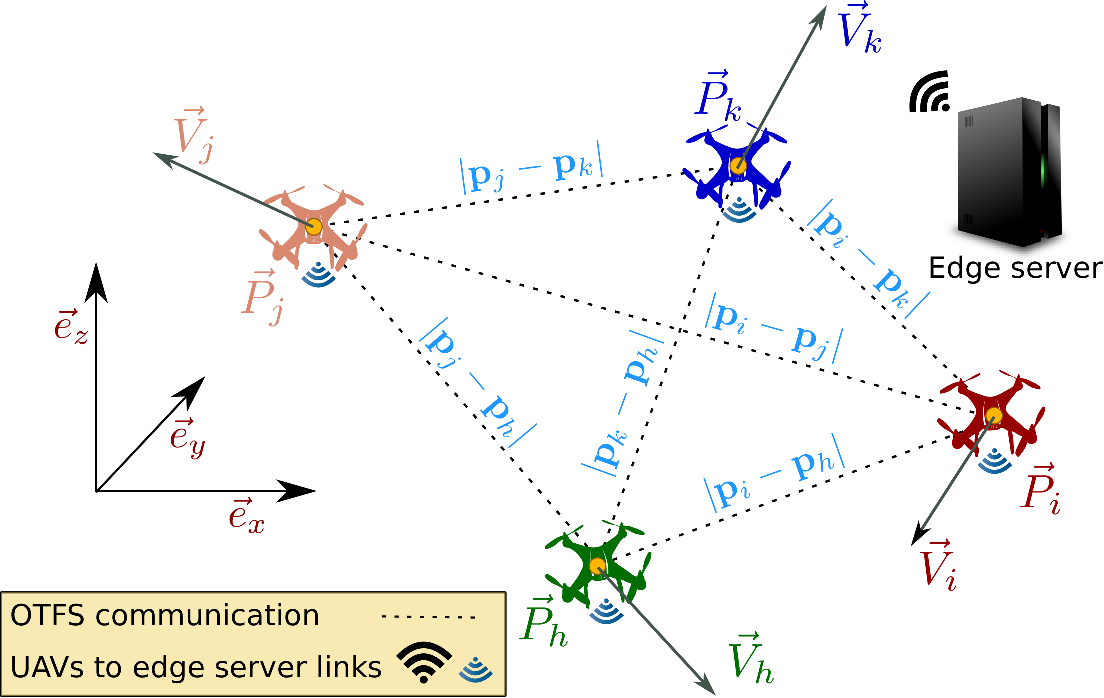}
\caption{Communicating UAVs assisted by an edge server. The UAV $u$  ($u{\in}\{h,i,j,k\}$)
  has position $\vec{P}_u$ and velocity $\vec{V}_u$, measured with
  respect to a common Cartesian coordinate system.  The notation
  $|\pv_u-\pv_v|$ denotes the distance between UAV $u$ and UAV $v$,
  where $\pv_u$ and $\pv_v$ are the components of the geometric
  vectors $\vec{P}_u$ and $\vec{P}_v$, respectively.}
\label{fig:UAV}
\end{figure}


\ematextz{In this work, we focus on a joint {\em network} communication and sensing scenario (shown in  Fig.~\ref{fig:UAV}) involving a UAV swarm, where nodes
communicate with each other as well as with an edge server via orthogonal time frequency space (OTFS) signaling.
OTFS-based communication takes advantage of the sparse multipath representation of the channels in the delay-Doppler domain, which can also be effectively used to acquire information about the range and velocity of the scatterers in the scene \cite{delay_Doppler_book}. }

{\color{black} Importantly, our solution exploits the delay-Doppler domain channel estimation used by the OTFS communication system to achieve 
precise positions and velocities of all UAVs within the swarm.
It is worth noting that, by operating in the delay-Doppler domain, OTFS can accurately resolve all the multiple paths with distinct delays and Doppler shifts.  
Indeed, although the path delay resolution used for position estimation is the same for both OTFS and orthogonal frequency-division multiplexing (OFDM)  (being inversely proportional to 
the waveform bandwidth),  OFDM cannot resolve paths with the same delay and distinct Doppler shifts, as it operates in the time-frequency domain. It follows that using OFDM results in large errors in the velocity estimation \cite{ema-radar}. 
}

\ematextz{Furthermore, unlike most of the positioning algorithms available in the
literature, our approach
is designed to use the relative echo delay (RED) measurements
between line-of-sight (LoS) and non-LoS (nLoS) paths for each OTFS communication link. 
{\color{black}The main advantage of using RED profiles over the more commonly assumed time-of-arrival (ToA) method is that, by exploiting nLoS paths,  it enables localization without the need for tight synchronization among the nodes.}  
The RED profile between LoS and nLoS paths is extracted from the delay domain enabling the UAVs to
operate without the need for fine-grained,  sample-level synchronization
(within the inverse of the communication bandwidth).
Specifically, the RED and Doppler shift of each path obtained from the delay-Doppler domain channel estimate enable us to derive the corresponding UAVs' positions and velocities.
{\color{black}
Even more importantly, RED enables us to take full advantage of the multipath channels in the network, which usually hurt the classic position  and velocity estimation techniques.}
}

Although the limited communication channel bandwidth yields 
low-resolution estimates of the REDs in the multipath channels between UAV pairs,  this is tempered by the availability at the edge server of the full multipath channel measurements.
The edge server first attempts to associate the RED measurements with the paths identified by the reflecting UAVs, i.e., to solve what in the localization literature is known as the {\em assignment problem.}
The edge server subsequently uses this information to estimate the UAVs'
positions, by applying a gradient descent algorithm that converges to the actual values of the UAVs' positions. 
As the last step, the UAVs' velocities can be derived from the Doppler shift measurements.

\ematextz{To jointly solve our specific assignment problem for all the RED profiles collected at the edge server, we employ a belief propagation (BP) algorithm that exploits the redundant RED measurements at all UAVs and their interdependence. 
Then, we propose a low complexity {\em turbo-iterative positioning} (TIP) algorithm that iteratively refines the position estimates by correcting residual errors in the initial solution to the assignment problem.}

\aletext{
To evaluate the performance of our approach, we resort to comparisons to the Cram\'er Rao bounds on the estimation errors.}
Moreover, by leveraging real-world traces, we
demonstrate that TIP (i) yields highly accurate estimated positions and velocities, and (ii) is   resilient to noise, approaching  the Cram\'er-Rao bound.

\ematextz{To summarize, our main contributions are as follows:
\begin{itemize}
    \item We define  the joint network communication and sensing scenario for a swarm of UAVs and propose the use of OTFS communications pilots for delay-Doppler channel estimation to acquire the position and velocity of all UAVs in the swarm from the RED profiles; 
    \item 
    We introduce a novel solution to the initial network RED assignment problem leveraging belief propagation. {\color{black} Importantly, our approach innovates beyond the existing literature where rarely  multipath propagation has been considered to help in classic positioning techniques; }
    
    \item 
    We design a new low-complexity iterative algorithm (TIP) to refine the initial assignment estimates and improve the accuracy of positions and velocities;  
    \item We demonstrate the effectiveness of our solution framework through numerical results and real-world UAV mobility traces, relative to the Cram\'er-Rao bound. 
    Importantly, we show that our solution substantially enhances the positioning accuracy beyond the limits of the communication bandwidth constraints.  
\end{itemize}
}

\aletext{The rest of the paper is organized as follows. We introduce the notation used in this work in Sec.\,\ref{sec:notation}. Then, after detailing the
assumptions and  the problem formulation, Sec.\,\ref{sec:system-problem}
 provides a roadmap to our
solution. 
The fundamental blocks of the proposed algorithm are  presented in
Secs.\,\ref{sec:maps}--\ref{sec:turbo}, with  
Sec.\,\ref{sec:maps} describing how to map the channel delay profiles to
the UAV identities using BP, and Sec.\,\ref{sec:positions} introducing the proposed estimators for the UAV's
positions and velocities.}   Sec.\,\ref{sec:turbo}
combines the previous results in a highly efficient positioning algorithm, 
and Sec.\,\ref{sec:CRLB} presents the joint Cram\'er Rao lower bound on the
variance of the position-velocity estimator.
Finally, we  analyze the performance of our algorithm 
 in Sec.\,\ref{sec:results}.  Conclusions and future
research directions are discussed in Sec.\,\ref{sec:concl}.

\section{Related Work\label{sec:related-work}}
Despite the extensive research on using 
OFDM waveforms for this
purpose~\cite{OFDM-1,OFDM-2}, the challenge of achieving precise
localization in nLoS and high-mobility scenarios
persists~\cite{emadi2023}. To address this, OTFS has been explored as an alternative to OFDM. 
In particular,~\cite{Sciancalepore21} demonstrates the effectiveness of
OTFS-modulated physical random access channel (PRACH) transmissions
for ToA based ranging, while~\cite{emadi2023} designs an
OFTS transceiver for transmitting and receiving positioning
reference signals and using ToA for belief propagation-based cooperative positioning.

{\color{black} In cooperative positioning techniques, a number of nodes (called agents) exchange information to enhance the overall accuracy of range and velocity estimation for all the nodes in the network~\cite{Wymeersch2009,Etzlinger2017}.
A large body of work foresees agents that combine information from multiple sensors such as cameras, LiDAR, GPS and inertial measurements units (IMU) to improve localization accuracy, see, e.g.,~\cite{Olson2020,Lu2021,Borelle2023,Wang2024}.
Alternatively, a localization approach based on antenna arrays and angle of arrival estimation of multipath signals is proposed in~\cite{Tensorization2022}.
A vast literature is available on multi-agent localization and tracking, for which we refer the interested reader  to~\cite{Wielandner2023}.
For example, the problem of localizing multiple targets in a multipath environment using a set of distributed single-antenna receivers is addressed  in~\cite{Aditya2018} and  solved using a Bayesian estimation technique. The same problem is  considered in~\cite{Jiang2007} using, as observations, a sequence of video frames.
Alternative approaches to multitarget tracking in clutter use probabilistic data association techniques \cite{Kiru2004,He2017}. 
We underline that all these localization techniques require additional sensing equipment or antennas. Instead, at the essence of joint communication and sensing is realizing localization through the  exclusive use of the communication system without spending additional resources~\cite{Liu2022,Sun2024}. 
}

Given the observations collected by the agents, the problem of determining the positions of the targets from a set of measurements can be viewed as a least square error estimation problem, where the association of the measurements to the variables is unknown. Jointly solving these two problems to the optimum has a prohibitive complexity. Nevertheless, high-performance multi-target tracking solutions based on message passing belief propagation (BP) algorithms have proven effective for their low complexity and flexibility~\cite{Meyer2018}. For instance, in~\cite{Leitinger2019} BP is applied to a simultaneous localization and mapping problem, in a multipath propagation environment where paths are associated to virtual anchors in a probabilistic fashion.

{\color{black} To the best of our knowledge, there are no existing
works leveraging RED profiles. The RED profile technique is part of our novel contribution, and we have coined such terminology. The RED profile technique constructively exploits the multipath nature of the channels and avoids the need for precise synchronization among UAVs. Typical solutions rely on the LoS paths only and assume precise synchronization. Indeed, it is worth noting that our approach differs from TDoA-based positioning with multiple sensor nodes, where only LoS paths are considered and signal cross-correlations need to be evaluated. Additionally, TDoA still requires either strong synchronization among the sensors  \cite{ Meyer2017},  or complex compensation techniques \cite{Wang2013}. }

Finally, we mention that an early version of this work has appeared in our conference paper \cite{NoiGlobecom2023}. With respect to \cite{NoiGlobecom2023}, we now integrate both positions and velocities estimation, and present a substantially improved version of the envisioned methodology as well as additional performance results.

\section{Notations \label{sec:notation}} 
We denote {\em geometric vectors}  by
  capital letters with an arrow on top, such as $\vec{A}$. 
  Boldface uppercase and lowercase letters denote matrices and vectors,
  respectively. The transpose of matrix $\Am$ is denoted by $\Am\Tran$,
  whereas $[\Am]_{i,j}$ indicates its $(i,j)$-th element. $ \Id$ is
  the identity matrix and the L2-norm of the vector $\av$ is
  represented by $|\av|$. 
 Given a Cartesian coordinate system with basis vectors
$\vec{e}_x$, $\vec{e}_y$, and $\vec{e}_z$,
 the geometric vector $\vec{A}$
  is represented as a three-dimensional column vector $\av=[\langle\vec{A},\vec{e}_x \rangle, \langle\vec{A},\vec{e}_y \rangle, \langle\vec{A},\vec{e}_z \rangle]\Tran$.  Sets, lists, or maps
  are denoted by calligraphic or Greek capital letters. \aletext{The notation $[\cdot]_{\neq}$
  means that all elements of the list $[\cdot]$ have different values.}
  The estimate of the quantity $a$ is denoted by $\widehat{a}$. The probability of
  an event $A$ is referred to as $\PP(A)$, while the probability
  density function (pdf) of the random variable $a$ is denoted by
  $f_a(a)$. 
  The symbol $\EE_a[\cdot]$ stands for expectation
  with respect to the random variable $a$. Finally, the Gaussian distribution
  with zero mean and variance $\sigma^2$ is denoted by
  $\Nc(0,\sigma^2)$ and the uniform distribution between $a$ and $b$ is denoted by $\Uc[a,b]$.
  
The main variables and parameters used throughout the paper are summarized in Table\,\ref{tab:parameters}.

\section{System Model and Problem Statement\label{sec:system-problem}}


The scenario under study includes  a swarm of $N$ UAVs located in a 3D flight area.  The position and the
velocity of the generic UAV $i$ ($i{=}1,{\ldots}, N$) are denoted by the
geometric vectors $\vec{P}_i$ and $\vec{V}_i$, respectively, which
refer to a common Cartesian coordinate system.  We also use the symbol
$\pv_i{=}[p_{i,x},p_{i,y},p_{i,z}]\Tran$ to denote the components of
$\vec{P}_i$ along the $x$, $y$, and $z$ axis of the coordinate system
and, similarly, $\vv_i{=}[v_{i,x},v_{i,y},v_{i,z}]\Tran$ to denote the
components of the velocity vector $\vec{V}_i$.  

A subset of $A$ UAVs act as {\em anchors}
for the system, i.e., their positions and velocities are perfectly known while the positions and velocities of
the remaining $\bar{N}{=}N{-}A$ UAVs are unknown and have to be estimated.
{\color{black} The role of the anchors is to resolve translation and rotation ambiguities when 
estimating the position and velocity of the UAVs. The use of anchors is commonly considered in the literature, e.g., in~\cite{Wielandner2023, Leitinger2019} and  their position is known to the edge server.
In a 3D space, at least 4 non-coplanar anchors are required. With only 3 anchors, the estimation problem becomes geometrically ill-conditioned and may result in invalid solutions. Since 3 anchors lie on the same plane, they are not sufficient to determine on which side of the plane a UAV is located.
}

As depicted in Fig.~\ref{fig:UAV}, the UAVs communicate with each other,
as well as with an edge server controlling the geographical area of
interest, using  OTFS waveforms.  The wireless channel connecting any
two UAVs presents high-mobility multipath fading characteristics. 
For simplicity,
we assume that no physical obstacle, other than the UAVs belonging to
the swarm, can reflect or block the signal. Thus, each
communication channel between any pair of UAVs is characterized by the
LoS path and $N{-}2$ nLoS paths due to a
single reflection from the other UAVs in the swarm.
Then the time-delay expression of the channel connecting UAVs
  $i$ and $j$ ($j{\neq} i$) is given by~\cite{delay_Doppler_book}:
\begin{equation} \label{eq:channel} h_{i,j}(t,\tau) = \sum_{k=1,k\neq
i}^N c_{i,j,k}\ee^{\jj 2\pi \nu_{i,j,k}t}\delta(\tau-\tau_{i,j,k})
\end{equation}
where $c_{i,j,k}$,  $\tau_{i,j,k}$, and  $\nu_{i,j,k}$ are, respectively, the channel coefficients,
the path delay, and the path Doppler shifts generated by the signal reflection on the $k$-th UAV. When $k{=}j$, we have the {\em LoS} path, otherwise we have the {\em nLoS} paths. We assume that the pilot power is sufficient for the channel estimation units of all OTFS receivers to identify all the channel parameters in (\ref{eq:channel}).

In a practical scenario, the UAV receivers are not
synchronized with respect to a common time reference. {\color{black} Leveraging on the delay-Doppler domain channel estimation of OTFS, 
the RED profiles of each communication link between UAVs $i$ and $j$ can be measured,  locally at each receiver. This is obtained by  using the delays of the echos from all the nLoS paths relative to the LoS path (zero relative delay).}
Then, as shown in Fig.\,\ref{fig:UAV}, the RED
$\tau_{i,j,k}$ corresponds to
the relative delay between the $k$-th nLoS path and the LoS path, and it is given by

\begin{equation}
  \tau_{i,j,k}=\tau_{j,k}+\tau_{k,i}-\tau_{i,j}
\end{equation}
where $\tau_{u,v}{=}\frac{1}{c}|\pv_u-\pv_v|$, $u,v{\in}\{i,j,k\}$, is the \ematextz{absolute} signal propagation delay
from UAV $u$ to UAV $v$, and $c$ is the speed of light. 
\ematextz{Clearly, for the LoS path,  $\tau_{i,j,j}{=}0$ by definition.}

The REDs $\tau_{i,j,k}$  depend upon the swarm
geometry and, for each one of them, we define the corresponding distance\footnote{We refer to this as the distance for convenience; more precisely, it is the difference in length between nLos and LoS paths.}
\begin{equation}\label{eq:delta_ijk} \delta_{i,j,k} \triangleq c\,
\tau_{i,j,k}= |\pv_j-\pv_k| + |\pv_k-\pv_i| -
|\pv_i-\pv_j|\,.
\end{equation} 

\begin{table}
\begin{center}
\caption{Variables and parameters
}
\label{tab:parameters}
\renewcommand{\arraystretch}{1.1}
\begin{tabular}{|p{0.15\columnwidth}|p{0.65\columnwidth}|}
\hline
\multicolumn{1}{|c|}{\bf Symbol} & \multicolumn{1}{|c|}{\bf Description}\\ \hline 
 $N$ & Total number of UAVs \\ \hline
 $A$ & Number of anchor UAVs\\ \hline
 $f_c$, $B$  & Signal central frequency and bandwidth\\ \hline
$T_f$  & OTFS frame duration\\ \hline
$M_d, N_D$  & Number of delay and Doppler bins\\ \hline
 $\pv_i$, $\vv_i$  &  Components of the position and velocity vectors of the $i$-th UAV\\ \hline  
 $\tau_{i,j,k}$ & RED between the $k$-th non-LoS path and the LoS path, when UAV $j$ communicates with UAV $i$\\ \hline
 $\delta_{i,j,k}$ & Distance corresponding to the propagation delay $\tau_{i,j,k}$\\ \hline
 $\widetilde{\delta}_{i,j,k}$ & Discretized distance $\delta_{i,j,k}$\\ \hline
 $\nu_{i,j,k}$ & Doppler shift of the signal transmitted by UAV $j$, reflected by UAV $k$, and received at UAV $i$ \\ \hline
 $\omega_{i,j,k}$ & Velocity corresponding to the Doppler shift $\nu_{i,j,k}$\\ \hline
 $\widetilde{\omega}_{i,j,k}$ & Discretized velocity  $\omega_{i,j,k}$\\ \hline
$\Dc_{i,j}$, $\Vc_{i,j}$ & Lists describing the profile of the channel connecting UAV $i$ and UAV $j$\\ \hline
$\muL_{i,j}$ & Map associating the elements of $\Dc_{i,j}$ and $\Vc_{i,j}$ to the UAV's identities \\ \hline

\end{tabular}
\end{center}
\end{table}

\begin{figure*}[ht]
  \centering
\includegraphics[width=0.7\textwidth]{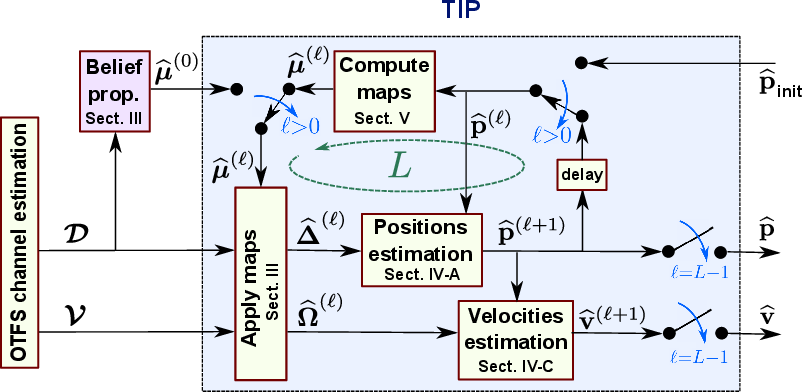}
\caption{\aletext{Scheme of the cold start algorithmic framework. An initial estimate of the maps, $\widehat{\muvL}^{(0)}$, is provided by the BP algorithm and fed to TIP. 
After $L$ iterations, TIP provides an estimate of the UAVs' positions, $\widehat{\pv}$, and velocities, $\widehat{\vv}$.}}
\label{fig:cold_start}
\end{figure*}

\begin{figure}[t]
  \centering
\includegraphics[width=0.7\columnwidth]{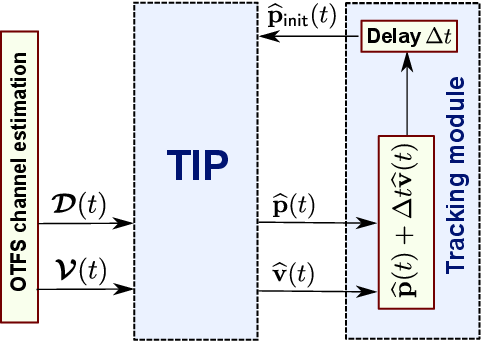}
\caption{\aletext{Scheme of the tracking algorithmic framework. At every time step $\Delta t$, TIP is provided with a new set of channel observations $\Dcb$ and $\Vcb$. TIP also takes as input an initial estimate of the UAV's positions, obtained at the previous time step and outputs the current estimates $\widehat{\pv}(t)$ and $\widehat{\vv}(t)$.}}
\label{fig:tracking}
\end{figure}

Notice also that, due to channel reciprocity,
$\delta_{i,j,k}{=}\delta_{j,i,k}$ $\forall i,j$, and $k{\neq}\{i,j\}$.
The delays $\tau_{i,j,k}$ or, equivalently, the distances
in~\eqref{eq:delta_ijk} can be measured through pilot-based OTFS channel
estimation techniques to a degree of accuracy dependent upon the signal
bandwidth $B$ \cite{OTFS_ChEst_Ravi}.

The Doppler shifts $\nu_{i,j,k}$ are instead functions of both UAV's
velocities and positions. For each of them, we define the
corresponding {\color{black}{\em radial velocity}}\footnote{{\color{black}The velocity $\omega_{i,j,k}$ is the radial component of the velocity vector $\vec{V}_j$, measured by the $i$-th receiving UAV  through the reflected path by the $k$-th UAV.}}, $\omega_{i,j,k}{\triangleq}
\frac{c}{f_c}\nu_{i,j,k}$, which
as shown in Appendix~\ref{app:velocity}, 
can be written as: 
\begin{equation}
\omega_{i,j,k}= (\vv_j-\vv_k)\Tran \uv_{j,k}(\pv)+(\vv_k-\vv_i)\Tran
\uv_{k,i}(\pv)  \label{eq:v_ijk}
\end{equation}
where $\pv{=}[\pv_1\Tran,{\ldots},\pv_N\Tran]\Tran$ is a column vector of size $3N$
stacking all components of the position vectors $\vec{P}_i$
($i{=}1,{\ldots}, N$), $f_c$ is the carrier frequency and the elements
of
\begin{equation}
\uv_{u,v}(\pv) = \frac{\pv_u-\pv_v}{|\pv_u-\pv_v|}\label{eq:versors}
\end{equation}
are the components of the versor pointing from UAV $v$ to UAV $u$.

\subsection{OTFS communication and channel estimation\label{sec:OTFS}}
{\color{black} OTFS is a communication scheme suitable for high-mobility communication environments (like the one considered in this work) exhibiting channels with multiple paths with different delays and Doppler shifts.
The goal of our paper is to demonstrate that accurate sensing of the position and velocity of highly mobile nodes in a network is possible by leveraging the OTFS scheme.  
Given the large diversity gains offered by OTFS, a reasonable SNR -- far below the one offered by conventional OFDM -- 
is sufficient to achieve a target error performance. 
Since the focus of this work is on localization and tracking, we do not report the performance of the communication links for the sake of brevity. 
However, for interested readers, this can be found in the vast literature demonstrating the superiority of OTFS over OFDM in a 
high-mobility scenario~\cite{delay_Doppler_book}.}

 In OTFS, the delay Doppler (DD) domain is discretized into an
$M_d{\times} N_D$ grid resulting in a delay resolution of $\Delta\tau
{=} \frac{T}{M_d}{=}\frac{1}{B}$ and a Doppler shift resolution of
$\Delta\nu{=} \frac{1}{N_D T}$, where $T{=}\frac{1}{\Delta f}$ is the
duration of a block in a frame of duration $T_f{=}N_D T$ and
$B{=}M_d\Delta f$ is the communication channel bandwidth
\cite{delay_Doppler_book}.  Simple channel estimation techniques, such
as the one in \cite{OTFS_ChEst_Ravi}, enable estimating the delay and
Doppler shift of each path with the above resolution.  
{\color{black} Pilot symbols can be embedded in the delay-Doppler together with the data, as shown in  \cite[Ch. 7]{delay_Doppler_book}. 
Such pilots are spread across the entire transmission frame ($T_f{=}NT$) and bandwidth ($B{=}M\Delta f{=}M/T$), 
timely capturing all the channel information relevant for the frame detection. By operating within $T_f$ seconds, we avoid the problem of channel information aging.}

By partitioning
the DD domain grid into disjoint rectangular tiles, UAVs can
simultaneously broadcast a pilot in a single frame along with data, thus allowing all
the receivers to perform channel estimation simultaneously within
$T_f$ seconds \cite{delay_Doppler_book}.
{\color{black}
This scheme avoids pilot interference among transmitting UAVs.}

Here we assume that the
multipath channel exhibits fractional delays and Doppler shifts, and
that, for each delay, there is only one Doppler shift.
In turn, this allows estimating the distances 
$\delta_{i,j,k}$ and the velocities $\omega_{i,j,k}$ from the measurements
  \begin{equation}\label{eq:distance_noise}
\widetilde{\delta}_{i,j,k}=\delta_{i,j,k}+\eta_{i,j,k}
\end{equation} 
\begin{equation}\label{eq:velocity_noise}
\widetilde{\omega}_{i,j,k}=\omega_{i,j,k}+\zeta_{i,j,k}
\end{equation}
where $\eta_{i,j,k}$ and $\zeta_{i,j,k}$ are the errors
affecting, respectively, distance and velocity estimation. 
These errors are due to the quantization noise caused by the limited DD resolutions
$\Delta\tau$ and $\Delta\nu$ as well as the effect of thermal
noise. While the thermal noise component contributing to  $\eta_{i,j,k}$ and $\zeta_{i,j,k}$ can be reduced by increasing the pilot power~\cite{OTFS_ChEst_Ravi}, the
quantization noise can only be reduced at the greater cost of increasing the bandwidth $B$ and the frame length $T_f$.

{\color{black} In addition to quantization noise some other causes of degradation in the delay-Doppler estimates may be considered. The motion of the UAV propellers may introduce spurious contributions to the channel delay-Doppler profile. Also, despite the frame time is typically a small fraction of a second (e.g., 20\,ms), sudden UAV accelerations may degrade the estimate of the Doppler shifts. 
As an example, consider signals with carrier frequency $f_c{=}5$\,GHz and a typical OTFS frame duration $T_f{=}20$\,ms. Then, the quantization step in the Doppler domain is $1/T_f{=}50\,Hz$. A UAV accelerating at 50\,m/s$^2$ (a rather significant $5g$ acceleration) varies its speed by $\Delta_v{=}1$\,m/s in $T_f$ seconds. 
This speed variation, assumed to be radial with respect to the observer, would result in a variation in the Doppler shift $\Delta_v f_c/c {\approx} 16$\,Hz, which is well below the 50\,Hz quantization step and has a negligible impact. Hence, in the following, we assume that estimation errors are only due to quantization noise.
}

{\color{black}
The channel measurement from each UAV can be sent to the edge server using a dedicated control channel. 
For a swarm composed of $N$ UAVs, there are $N(N{-}1)$ multipath channels, each of them having $N-1$ paths with their delay and Doppler 
information.
Hence, the total number of measurements is  
$2N(N{-}1)^2$, which yields a negligible communication overhead\footnote{\color{black}For example, assuming each delay and Doppler measurement takes 2 bytes, 
the total number of bytes required for transferring all lists is $4N(N{-}1)^2$, which, for a swarm composed of $N{=}20$ UAVs 
adds up to about 30\,kbytes. If the updated lists are transferred every second to the edge server, the total required overhead is about 
240\,kbit/s which can be easily supported by the control channels of modern communication systems.}.
}

\subsection{Problem statement and roadmap}
If the UAV's positions are unknown, the $i$-th UAV cannot
associate each path in~\eqref{eq:channel} with the identity of the UAV
that generated it. Rather, it can only arrange the delays
$\widetilde{\delta}_{i,j,k}$ in the list
\begin{equation}\label{eq:Dc} 
\Dc_{i,j}=[d_{i,j,1}, \ldots, d_{i,j,N-1}]
\end{equation}
and the velocities $\widetilde{\omega}_{i,j,k}$ in the list
\begin{equation} \label{eq:Vc}
\Vc_{i,j}=[v_{i,j,1}, \ldots, v_{i,j,N-1}]
\end{equation}
where the elements of $\Dc_{i,j}$, as well as of 
$\Vc_{i,j}$, appear in increasing order (i.e.,
$d_{i,j,1}{\le} d_{i,j,2} {\le} {\ldots} {\le} d_{i,j,N-1}$). 
In general, $d_{i,j,m} {=} \widetilde{\delta}_{i,j,k}$, and $v_{i,j,m}
{=} \widetilde{\omega}_{i,j,k}$, if the elements at position $m$ in
both lists correspond to the DD signature of the $k$-th
UAV.  In other words, we can associate with each list $\Dc_{i,j}$ and
$\Vc_{i,j}$ a bijective map of the indices
\[ \mul_{i,j}: \{1,\ldots,N\}{\setminus} \{i\} {\to} \{1,\ldots,N-1\} \]
such that $\muL_{i,j}(k){=}m$. Note that $d_{i,j,1}{=}0$ corresponds
to the LoS path between UAV $j$ and UAV $i$, since it is the shortest
possible \ematextz{relative delay}.
By defining the set of all lists $\Dcb {\triangleq} \{\Dc_{i,j} |
i,j{=}1,\ldots,N,\,i{\neq} j\}$ and $\Vcb {\triangleq} \{\Vc_{i,j} |
i,j{=}1,\ldots,N,\, i{\neq} j\}$, and the set of all maps
$\muvL{\triangleq} \{\muL_{i,j} | i,j{=}1,\ldots,N,\, i{\neq} j\}$,
the problem addressed in this work can be stated as follows:

{\bf Problem statement} -- {\em Given $N$ UAVs communicating between each-other using OTFS through channels as in~\eqref{eq:channel}, and the lists of noisy delay-Doppler profiles $\Dcb$ and $\Vcb$, how can an edge
    server reliably estimate the UAVs' geometric positions and
    velocities vectors, $\vec{P}_i$ and $\vec{V}_i$, $i{=}1,{\ldots}, 
    N$?}

{\bf Solution roadmap.}  We answer this question by developing a new
  positioning algorithm called TIP (Turbo Iterative Positioning). 
\aletext{Using the RED and Doppler shift measurements of
  the OTFS channel estimation on all the UAVs links, collected in the
  set of lists $\Dcb$ and $\Vcb$ (resp.), TIP enables two operational modes:
  (i) {\em cold start} (Fig.~\ref{fig:cold_start}) where no prior information on positions and
  velocities is available, and (ii) {\em tracking} (Fig.~\ref{fig:tracking}) where the estimated
  positions and velocities are updated every $\Delta t$ seconds using
  the previous estimates.  The cold start mode is based on a BP  algorithm to produce an initial estimate of the maps, $\widehat{\muvL}^{(0)}$, assigning the identity of the scatterers in each path of each link. 
  
  The TIP algorithm takes as input such initial maps' estimate and iterates $L$ times to refine the estimated positions and velocities, and correct residual errors in the corresponding maps. At the core of the iterative loop of TIP, a gradient descent algorithm
  minimizes an error function to estimate the positions at step
  $\ell+1$, given the currently estimated positions and maps. The same
  maps are applied to the lists of Doppler shifts $\Vcb$ to estimate
  the velocities. 
  In the tracking mode, 
  the estimates of positions and
  velocities are combined to predict the next positions after $\Delta
  t$ seconds, which can be fed back to TIP as a reliable starting point for the gradient descent algorithm.}   
  
The overall approach is detailed in Sec.~\ref{sec:maps}--\ref{sec:turbo} and includes
  the following blocks:
\begin{enumerate}
\item {\em Belief Propagation}~--~provides an initial estimate of the
  maps $\muvL$, given the sets of lists $\Dcb$ and $\Vcb$. Towards this goal, we
  relax the set of deterministic maps $\muvL$ to \aletext{a set of random maps}
  and use a BP approach to obtain an estimation 
  $\widehat{\muvL}^{(0)}$
  (see Sec.~\ref{sec:maps}).
\item 
  \aletext{{\em Apply maps}~--~permutes the RED and Doppler shift lists $\Dcb$
  and $\Vcb$ according to the estimated maps $\widehat{\muvL}^{(\ell)}$, 
  to produce the permuted lists $\widehat{\Deltam}^{(\ell)}$ and
  $\widehat{\Omegam}^{(\ell)}$, respectively
  (see Sec.~\ref{sec:apply_maps}).} 
\item 
\aletext{{\em Positions estimation}~--~computes the new UAVs' position estimates
$\widehat{\pv}^{(\ell+1)}$, given the lists $\widehat{\Deltam}^{(\ell)}$ from {\em Apply
  maps} and $\widehat{\pv}^{(\ell)}$, using a gradient descent algorithm (see  Sec.~\ref{sec:gradient_descent}}).
\item 
\aletext{{\em Velocity estimation}~--~provides estimates $\widehat{\vv}^{(\ell+1)}$
  of the UAV's velocities using the estimates $\widehat{\pv}^{(\ell+1)}$, and
  the set of lists $\widehat{\Omegam}^{(\ell)}$ from {\em Apply maps}
  (see Sec.~\ref{sec:velocities}).}
\item \aletext{{\em Compute maps}~--~computes the set of maps $\widehat{\muvL}^{(\ell)}$
  corresponding to the current position estimates $\widehat{\pv}^{(\ell)}$ (see Sec.~\ref{sec:turbo}).}
\end{enumerate}


\section{Solving the network assignment problem through BP\label{sec:maps}}
To associate channel observations with UAVs' identities in each link, the edge server has to estimate the maps $\muvL$. This is an instance of an assignment problem that can be solved by using a probabilistic approach.
\aletextz{Due to the presence of noise
we can relax the set of deterministic maps $\muvL$ to a set of random maps
     $\boldsymbol{\Mc}{\triangleq} \{\Mc_{i,j}| i,j{=}1,\ldots,N,
     i\neq j\}$ having joint posterior distribution $\PP\left(\boldsymbol{\Mc}\right |\Dcb,\Vcb)$, given the channel measurements $\Dcb$ and $\Vcb$.  
     The optimal estimation of the maps would require to solve 
\begin{equation}
    \label{eq:argmax_mu}
\widehat{\muvl} = \arg\max_{\muvL}
\PP\left(\boldsymbol{\Mc}=\muvl\right |\Dcb,\Vcb)
\end{equation}
which has an intractable complexity of $O([(N{-}1)!]^{N(N-1)})$.}
To overcome this complexity, we propose a BP approach 
based on a factor graph with specific variable nodes and check nodes~\cite{Meyer2018}. In the rest of this section, we describe how to specialize this proposed approach to our setting. 
\aletextz{To simplify the presentation, we first restrict BP to the case where only the channel measurements $\Dcb$ are used.
Then  we extend this  to the case where both $\Dcb$ and $\Vcb$ are employed in~\eqref{eq:argmax_mu}.}

\subsection{Using check node relations on delays}
Let us first focus on a
subset of four UAVs, respectively labelled by $i,j,k,h$, as depicted
in Fig.~\ref{fig:UAV}. 
The figure also highlights the actual distances
$|\pv_u-\pv_v|$ between the UAVs, for $u,v{\in} \{i,j,k,h\}$.  
By recalling the expression of $\delta_{i,j,k}$ in~\eqref{eq:delta_ijk},
it is easy to observe that for all quadruples of UAVs in the swarm, we have:
\begin{equation}\label{eq:4node_equation}
  \delta_{i,j,k} - \delta_{i,j,h} +\delta_{i,k,h}-\delta_{j,h,k} = 0
\end{equation}
where $i,j,k,h {\in} \{1,\ldots,N$\}, $j{\neq}i$, $k{\neq} \{i,j\}$,
$h{\neq}\{i,j,k\}$.  This implies that one can find the correct
associations $\muL_{i,j}$ between the entries of the list $\Dc_{i,j}$
and the UAVs' identities by searching through the elements of lists
$\Dc_{i,j}$, $\Dc_{i,k}$, and $\Dc_{j,h}$, until a relation such as the
one in~\eqref{eq:4node_equation} holds.  More precisely, given the
ordered lists $\Dc_{i,j}$, $\Dc_{i,k}$, and $\Dc_{j,h}$, if for some
integers $m,n,s,t{\in}\{1,\ldots,N-1\}$ and such that $m{\neq} n$, the following relation holds in the absence of noise: 
\begin{equation}
\label{eq:4node_equation_2}
  d_{i,j,m} - d_{i,j,n} +d_{i,k,s}-d_{j,h,t} = 0\,,
\end{equation}
then, by
comparing~\eqref{eq:4node_equation} to~\eqref{eq:4node_equation_2}, we
can deduce the maps $\muL_{i,j}(k){=}m$, $\muL_{i,j}(h){=}n$,
$\muL_{i,k}(h){=}s$, and $\muL_{j,h}(k){=}t$.  

It is worth  noting that there are
$O(N^8)$ constraint equations like \eqref{eq:4node_equation_2} that
need to be satisfied simultaneously by the $d$'s in all the lists
$\Dcb$.
Additionally, whenever the distances
$\delta_{i,j,k}$ are affected by noise, the
left-hand side of~\eqref{eq:4node_equation_2} is, in general, not zero but random.  

\subsection{Defining the variable nodes and computing their marginals}
 We observe that 
the random map $\Mc_{i,j}$ can be represented by a set of $N{-}1$ correlated discrete random variables $\Mc_{i,j,k}$,
     $k{=}\{1,{\ldots},N\}/\{i\}$ having support on the integer set $\{1,{\ldots},N-1\}$ 
     and probability mass function (pmf) $\pi_{i,j,k,m} {\triangleq}
     \PP(\Mc_{i,j,k}{=}m | \boldsymbol{\Dc})$ with
     $\sum_{m=1}^{N-1}\pi_{i,j,k,m}{=}1$.
These correspond to the variable nodes in the factor graph.
  The terms $\pi_{i,j,k,m}$ are the marginals of the joint posterior distribution $\PP\left(\boldsymbol{\Mc}\right |\Dcb)$
   and represent the probability that the RED profile of the $k$-th
   UAV is identified by the $m$-th element of the list $\Dc_{i,j}$. 
In the absence of quantization noise, $\eta_{i,j,k}$, the pmf of
$\Mc_{i,j,k}$ coincides with a pmf with probability 1 in correspondence with the actual value $\muL_{i,j}(k)$, i.e., the correct identity of the UAV.
We first show how to
compute such marginals using a BP approach, and then we describe how
to estimate the maps $\muvL$, given the marginals $\pi_{i,j,k,m}$ in Sec.~\ref{sec:estimating_maps}.

To compute the marginals $\pi_{i,j,k,m}$, 
we first need to consider some special cases
of~\eqref{eq:4node_equation_2}, which must be treated
separately. Specifically, we observe that, for every $i{\neq} j$, when $m{=}1$, the first
term of~\eqref{eq:4node_equation_2} is $d_{i,j,1}{=}0$ by definition 
since it is the smallest possible RED and, thus, it refers to the LoS
path\footnote{{\color{black} Because of quantization noise, some entries in the list 
  $\mathcal{D}_{i,j}$ might take the same value. This means that, by observing the list, the UAVs generating such entries are indistinguishable.
  We also know that one of the elements of the list is due to the LoS path associated with zero delay.
  If two or more entries of $\mathcal{D}_{i,j}$ take on the same value due to quantization, we arbitrarily associate one of them with the LoS path. We then rely on the TIP iterations to correct any error in the choice between the two.}}, i.e., the path
connecting node $j$ with node $i$. It follows that
$\muL_{i,j}(j){=}1$, which also implies $k{=}j$.  In this case,
\eqref{eq:4node_equation} becomes meaningless since its left-hand side is
identically zero. Similarly, referring to~\eqref{eq:4node_equation_2}, we deduce $\muL_{i,k}(k){=}1$ when
$s{=}1$ and $\muL_{j,h}(h){=}1$, and $t{=}1$.  Summarizing, for all
$i{\neq} j$, we can handle these special cases by setting
$\pi_{i,j,k,1}{=}1$ for $k{=}j$ and $\pi_{i,j,k,1}{=}0$ for $k{\neq} j$.

Specializing the BP approach to the notation in~\cite{Meyer2018}, we 
only consider the indices $m,n,s,t$, with $m{\neq} n$ ranging in
$\{2,{\ldots},N{-}1\}$ and factorize the joint posterior
pmf $\PP(\boldsymbol{\Mc} | \boldsymbol{\Dc})$
as
\begin{equation}  \PP(\boldsymbol{\Mc} | \boldsymbol{\Dc}) = \prod_{Q \in \Qc }\psi_{Q}\left(\boldsymbol{\Mc}_Q|\boldsymbol{\Dc}_Q\right) \label{eq:factorization}
  \end{equation}
where
\begin{equation} \Qc = \{ [i,j,k,h] | [i,j,k,h] \in \{1,\ldots,N\}^4, [i,j,k,h]_{\neq}\} 
\label{eq:quadruple}
\end{equation} 
is the set of all possible quadruples $Q$ of distinct UAVs. For a given $Q{=}[i,j,k,h]$, 
$\boldsymbol{\Mc}_Q {\subseteq} \boldsymbol{\Mc}$ denotes the set of random variables
$\boldsymbol{\Mc}_Q {=}\{\Mc_{i,j,k}, \Mc_{i,j,h},
\Mc_{i,k,h},\Mc_{j,h,k}\}$, and
$\boldsymbol{\Dc}_Q{=}\{\Dc_{i,j},\Dc_{i,k}, \Dc_{j,h}\}$.  In analogy
with~\eqref{eq:4node_equation_2}, the factor
$\psi_{Q}(\boldsymbol{\Mc}_Q|\boldsymbol{\Dc}_Q)$ is associated with the
equation
\begin{equation}
\label{eq:check_node_equations}
  d_{i,j,\Mc_{i,j,k}} - d_{i,j,\Mc_{i,j,h}}+d_{i,k,\Mc_{i,k,h}} - d_{j,h,\Mc_{j,h,k}} = 0\,, 
\end{equation}
which depends on the subset of random variables $\boldsymbol{\Mc}_Q$
and is parameterized by $\boldsymbol{\Dc}_Q$. In practice, when the
random variables in $\boldsymbol{\Mc}_Q$ take on the specific values
$\Mc_{i,j,k}{=}m$, $\Mc_{i,j,h}{=}n$,
$\Mc_{i,k,h}{=}s$, $\Mc_{i,j,k}{=}t$, ~\eqref{eq:check_node_equations} converts
to~\eqref{eq:4node_equation_2} and the function
$\psi_{Q}\left(\Mc_{i,j,k}{=}m, \Mc_{i,j,h}{=}n,
\Mc_{i,k,h}{=}s,\Mc_{i,j,k}{=}t|\boldsymbol{\Dc}_Q\right)$ provides
a measure of the likelihood that equation~\eqref{eq:4node_equation_2} is
satisfied. Note that the left-hand side of~\eqref{eq:4node_equation_2} contains the sum of 4 independent quantization errors characterized by the
uniform distribution $f_\eta(\cdot)$ with support in $[-\frac{c}{2B},\frac{c}{2B}]$.
Therefore, the distribution of such sum is $g(\cdot){=}f_\eta(\cdot ) {*} f_\eta(\cdot ) {*} f_\eta(\cdot ){*}f_\eta(\cdot )$
where $*$ denotes the convolution operator. 
The above definition of $g(\cdot)$ can be explained by the fact
that~\eqref{eq:4node_equation_2} is a sum of four quantized values of actual
REDs, which are unknown, but within the quantization intervals. Due to the rounding operation, the quantization error is deterministically
dependent upon the actual value of a RED, but the uncertainty of such a
value can be translated onto the uniform distribution $f_{\eta}(\cdot)$ centered around
the center point of a quantization interval for the $\eta$'s to
represent a random quantization noise.
Summarizing, for any $Q=[i,j,k,h]$ and integers $m,n,s,t {\in} \{2,\ldots,N-1\}$, $m\neq n$,
let $z_Q^{m,n,s,t}$ be the l.h.s. of~\eqref{eq:4node_equation_2}. Then, we can write
\[ \psi_Q\left(\boldsymbol{\Mc}_Q=[m,n,s,t]|\boldsymbol{\Dc}_Q\right) =g\left(z_Q^{m,n,s,t}\right)\,.\]

For convenience, let us define the set of all possible triples of distinct UAVs
\[ \Tc = \{ [i,j,k] | [i,j,k] \in \{1,\ldots,N\}^3, [i,j,k]_{\neq}\} \]
so that each triple $T{\in} \Tc$ uniquely identifies the random variable
$\Mc_T$.  Then, the factorization in~\eqref{eq:factorization} can be
represented by a {\em factor graph} whose {\em variable nodes} are the
random variables $\Mc_T$, $T\in \Tc$ and the {\em factor nodes} or
{\em check nodes} are the functions $\psi_Q(\cdot)$, for all $Q\in
\Qc$. In such graph each check node $Q\in \Qc$ is connected to 4 
variable nodes. Specifically, the check node $Q{=}[i,j,k,h]$ has neighbours $T_1{=}[i,j,k]$, $T_2{=}[i,j,h]$, $T_3{=}[i,k,h]$, and
$T_4{=}[j,h,k]$.  Each variable node $T{=}[i,j,k] {\in} \Tc$ has $4(N{-}3)$
neighbouring check nodes, i.e., $[i,j,k,\ell]$, $[i,j,\ell,k]$,
$[i,\ell,j,k]$, and $[\ell,i,j,k]$, for $\ell{\in}
\{1,{\ldots},N\}{\setminus} \{i,j,k\}$. The graph has cycles of a minimum length of 4, 
since both check nodes $Q_1{=}[i,j,k,h]$ and $Q_2=[i,j,h,k]$ are connected to the variable nodes $T_1{=}[i,j,k]$ and $T_2{=}[i,j,h]$. Therefore, convergence of the BP algorithm cannot be theoretically guaranteed.

Let $\Nc_2$ be the set $\{2,{\ldots},N-1\}$. Then, in the BP algorithm, the
messages flowing from the check node $Q$ to the adjacent variable
nodes $T_1$, $T_2$, $T_3$, and $T_4$ are given by
\begin{eqnarray*}
  \zeta_{Q\to T_1}(m) &\hspace{-2.7ex}{=}& \hspace{-4ex}\sum_{\substack{n,s,t\in \Nc_2 \\ n\neq m}}\!g(z_Q^{m,n,s,t}) \lambda_{T_2\to Q}(n) \lambda_{T_3\to Q}(s) \lambda_{T_4\to Q}(t) \non
  \zeta_{Q\to T_2}(n) &\hspace{-2.7ex}{=}& \hspace{-4ex}\sum_{\substack{m,s,t\in \Nc_2 \\ m\neq n}}\!\!\!\!g(z_Q^{m,n,s,t}) \lambda_{T_1\to Q}(m) \lambda_{T_3\to Q}(s) \lambda_{T_4\to Q}(t) \non
\end{eqnarray*}
\begin{eqnarray}
  \zeta_{Q\to T_3}(s) &\hspace{-2.7ex}{=}& \hspace{-4ex}\sum_{\substack{n,m,t\in \Nc_2 \\ m\neq n}}\!\!\!\!\!g(z_Q^{m,n,s,t}) \lambda_{T_1\to Q}(m) \lambda_{T_2\to Q}(n) \lambda_{T_4\to Q}(t) \non
  \zeta_{Q\to T_4}(t) &\hspace{-2.7ex}{=}& \hspace{-4ex}\sum_{\substack{m,n,s\in \Nc_2 \\ m\neq n}}\!\!\!\!\!g(z_Q^{m,n,s,t}) \lambda_{T_1\to Q}(m) \lambda_{T_2\to Q}(n) \lambda_{T_3\to Q}(s)\,. \nonumber \label{eq:QtoT}\\
\end{eqnarray}
Let $\Qc_T{\subseteq} \Qc$ be the set of check nodes connected to the variable node $T$.
The message flowing from the variable node $T$ to check node $Q$ is given by
\begin{equation} \label{eq:TtoQ}
 \lambda_{T\to Q}(m)  = \prod_{Q'\in\Qc_T \setminus \{Q\}} \zeta_{Q'\to T}(m)
\end{equation}
for all $m{\in} \Nc_2$.  The procedure is iterative and is initialized
by setting $\lambda_{T\to Q}(m){=}1/(N{-}2)$, for $m{\in} \Nc_2$.
Note that the expressions in~\eqref{eq:QtoT} and~\eqref{eq:TtoQ} need to be normalized to add up to 1 to form a probability vector with $N{-}2$ entries.
At each iteration, the marginal probabilities at variable node $T{=}[i,j,k]$, i.e., the beliefs,
are computed as
\begin{equation} \label{eq:BPmarginals} \pi_{i,j,k,m}  = \frac{\prod_{Q\in\Qc_T} \zeta_{Q\to T}(m)}{\sum_{m=2}^{N-1}\prod_{Q\in\Qc_T} \zeta_{Q\to T}(m)}\,. \end{equation}
The above procedure is summarized in Algorithm~\ref{alg:belief}, which has complexity  $O(N^8)$.

\subsection{Estimating the maps $\muvL$ given the marginals $\pi_{i,j,k,m}$\label{sec:estimating_maps}}
To estimate the maps, we propose the heuristic greedy approach in Algorithm~\ref{alg:belief2}, which takes as input the
marginals $\pi_{i,j,k,m}$ obtained from  Algorithm~\ref{alg:belief} and has complexity $O\left(N(N{-}1)(N{-}2)\right)$.  

For each pair of UAVs, $(i,j)$, the algorithm collects the marginals
$\pi_{i,j,k,m}$ in the $N\times (N{-}1)$ matrix $\Pim$.  It then works
iteratively and, at each step, finds the most likely association in
the map. Specifically, it seeks the largest entry of $\Pim$, records
its row index $k'$ and its column index $m'$, and sets
$\widehat{\muL}_{i,j}(k'){=}m'$. Next, the algorithm sets the $k'$-th row and the
$m'$-th column of $\Pim$ to zero: this operation is necessary because
no other element of the map can be assigned to value $m'$ in the
following steps of the algorithm.  The procedure ends when all values
of the elements of $\Pim$ become zero, i.e., a decision is made on all
the elements of the map $\widehat{\muL}_{i,j}$.

\subsection{Extension using check nodes on Doppler shifts \label{sec:extension_Doppler}}
\aletextz{We now show how the measurements of  Doppler shifts can be added to the BP algorithm.
First, an additional set of check-nodes equations are added to those described by~\eqref{eq:4node_equation}. These new equations involve the Doppler shifts measurements, $\Vcb$, and can be derived by recalling the expression of $\omega_{i,j,k}$ in~\eqref{eq:v_ijk}. We have
\begin{equation}\label{eq:4node_equation_v}
  \omega_{i,j,k} + \omega_{i,j,h} -\omega_{k,h,i}-\omega_{k,h,j} = 0
\end{equation}
for every quadruple $Q{\in}\Qc$ defined in \eqref{eq:quadruple}. Using both~\eqref{eq:4node_equation} and~\eqref{eq:4node_equation_v}, 
the joint posterior probability mass function $\PP(\boldsymbol{\Mc} | \Dcb,\Vcb)$ can be factorized as
\begin{equation}  \PP(\boldsymbol{\Mc} | \Dcb,\Vcb) = \prod_{Q \in \Qc}\psi_{Q} \left(\boldsymbol{\Mc}_{Q}|\Dcb_{Q}\right) \phi_{Q} \left(\boldsymbol{\Mc}'_{Q}|\Vcb_{Q}\right) \label{eq:factorization2}
  \end{equation}
  where $\boldsymbol{\Mc}'_Q {\subseteq} \boldsymbol{\Mc}$ denotes the set of random variables
$\boldsymbol{\Mc}'_Q {=}\{\Mc_{i,j,k}, \Mc_{i,j,h},
\Mc_{k,h,i},\Mc_{k,h,j}\}$, and
$\Vcb_Q{=}\{\Vc_{i,j},\Vc_{k,h}\}$. Similarly to the discussion following ~\eqref{eq:check_node_equations}, we define
\[ \phi_Q\left(\boldsymbol{\Mc}'_Q=[m,n,s,t]|\Vcb_Q\right) =g'\left(w_Q^{m,n,s,t}\right)\]
where $g'(\cdot){=}f_\zeta(\cdot ) {*} f_\zeta(\cdot ) {*} f_\zeta(\cdot ){*}f_\zeta(\cdot )$
and $f_\zeta(\cdot)$ is a uniform distribution with support $[-\frac{c}{2T_f f_c},\frac{c}{2T_f f_c}]$. Furthermore, $w_Q^{m,n,s,t}$ is given by
\[ w_Q^{m,n,s,t} = v_{i,j,m} + v_{i,j,n} -v_{k,h,s}-v_{k,h,t}\]
for the integers $m,n,s,t {\in} \{1,{\ldots},N-1\}$ such that $m{\neq} n$ and $s{\neq} t$.
Without going into the details, 
\eqref{eq:QtoT}--\eqref{eq:BPmarginals} can then be modified to  account for the additional check nodes. Algorithms~\ref{alg:belief} and~\ref{alg:belief2} need to be modified accordingly.}

\subsection{Applying the estimated maps to lists $\Dcb$ and $\Vcb$\label{sec:apply_maps}}
Once the
  estimated map $\widehat{\muL}_{i,j}$ is available, it can be applied
  to lists $\Dc_{i,j}$ and $\Vc_{i,j}$ in order to associate to each
  reflecting UAV, $k$, an estimate of the distance
  $\delta_{i,j,k}$ and of the velocity $\omega_{i,j,k}$.
  We recall that the lists $\Dc_{i,j}$ and $\Vc_{i,j}$
  are defined in~\eqref{eq:Dc} and~\eqref{eq:Vc}, respectively.
  
  Specifically, the distance associated with the $k$-th path
  is estimated as $\widehat{\delta}_{i,j,k}{=} d_{i,j,\widehat{m}}$
  where $\widehat{m} {=} \widehat{\muL}_{i,j}(k)$, and $d_{i,j,\widehat{m}}$ is an element of
  the list $\Dc_{i,j}$.
  Similarly, the velocity associated with the $k$-th path
  is estimated as $\widehat{\omega}_{i,j,k}{=}v_{i,j,\widehat{m}}$
  where $v_{i,j,\widehat{m}}$ is an element of the list $\Vc_{i,j}$.
  This procedure is represented by the block labeled ``Apply maps'', 
  shown in Fig.~\ref{fig:cold_start}.

\begin{algorithm} 
\aletext{\caption{$\{\pi_{i,j,k,m}\}=$ {\rm ComputeMarginals}$(\boldsymbol{\Dc})$}\label{alg:belief}
\begin{algorithmic} 
  \REQUIRE $N$, $\Dcb$, $I_{\mu}{>}0$ 
  \FOR{$i,j{=}1,\ldots,N$,
    $j{\neq} i$} 
  \STATE $\pi_{i,j,k,1} {\gets} 1$, for $k{=}j$
  and $\pi_{i,j,k,1} {\gets} 0$ for $k{\neq} \{i,j\}$
  \ENDFOR
  \FOR{$T\in \Tc$}
    \FOR{$Q\in \Qc$}
      \FOR{$m{=}2,\ldots,N{-}1$} 
      \STATE initialize $\lambda_{T\to Q}(m) = \frac{1}{N{-}2}$
  \ENDFOR
  \ENDFOR
  \ENDFOR
  \FOR{$\ell{=}1,\ldots, I_\mu$} 
  \FOR{$Q\in \Qc$}
    \STATE let $\Tc_Q$ be the neighbours of $Q$
    \FOR{$T\in \Tc_Q$}
    \FOR{$b=2,\ldots N-1$}
    \STATE compute $\zeta_{Q\to T}(b)$ according to~\eqref{eq:QtoT}
    \ENDFOR
    \ENDFOR
  \ENDFOR
 \FOR{$T \in Tc$}
   \STATE let $\Qc_T$ be the neighbours of $T$
   \FOR{$Q \in \Qc_T$}
      \FOR{$b=2,\ldots N-1$}
        \STATE compute $\lambda_{T\to Q}(b)$ according to~\eqref{eq:TtoQ}
      \ENDFOR
   \ENDFOR
  \ENDFOR
  \ENDFOR
  \FOR{$T\in \Tc$}
    \STATE let $\Qc_T$ be the neighbours of $T$
    \STATE let $T=[i,j,k]$ 
    \FOR{$m=2,\ldots,N-1$}
      \STATE compute the marginals $\pi_{i,j,k,m}$ using~\eqref{eq:BPmarginals}
      \ENDFOR
      \ENDFOR
  \RETURN $\{\pi_{i,j,m,k}\}$
\end{algorithmic}} 
\end{algorithm}

\begin{algorithm} 
\aletext{\caption{$\widehat{\muvL}{=}$ {\rm
    EstimateMaps}$(\Dcb)$}\label{alg:belief2}
\begin{algorithmic} 
\REQUIRE $N$,$\Dcb$ \STATE $\{\pi_{i,j,k,m}\} {\gets} $ {\rm
  ComputeMarginals}$(\boldsymbol{\Dc})$ \FOR{$i,j=1,\ldots,N$,
  $i{\neq} j$} \STATE $[\Pim]_{k,m} {\gets}
\left\{\begin{array}{ll}\pi_{i,j,k,m} & k {\neq} i\\ 0 &
k{=}i \end{array}\right.$ \WHILE{$\Pim {\neq} \zerov$} \STATE $[k',m']
                 {\gets} \arg\max_{k,m} ([\Pim]_{k,m})$ \STATE
                 $\widehat{\muL}_{i,j}(k'){\gets} m'$ \STATE
                 $[\Pim]_{q,m'}{\gets} 0$ for $q{=}1,\ldots, N$ \STATE
                 $[\Pim]_{k',q} {\gets} 0$ for $q{=}1,\ldots,N{-}1$
                 \ENDWHILE \ENDFOR \RETURN $\widehat{\muvL}$
\end{algorithmic}} 
\end{algorithm}

\section{Estimating the Node Positions and velocities\label{sec:positions}}
Once the estimates $\widehat{\muvL}$ of the maps $\muvL$ have been used by the edge server to associate the elements of the lists $\boldsymbol{\Dc}$ with the UAVs that generated them (see Sec~\ref{sec:apply_maps}),
we apply the gradient descent (GD)
algorithm to minimize the square error, $\Ec$, between
$\widehat{\delta}_{i,j,k}$ and the tentative channel observations
$\theta_{i,j,k}$.
\aletext{We remark that, in the case of perfect maps
estimates, the edge server can retrieve the channel
observations $\widehat{\delta}_{i,j,k}{=}\widetilde{\delta}_{i,j,k}$ in their correct ordering.}

\subsection{\aletext{Gradient descent algorithm}\label{sec:gradient_descent}}
\ematextz{Without loss of generality, we describe the gradient descent algorithm under the assumption that the maps have been correctly estimated.} Let $\tv_i{=}[t_{i,x}, t_{i,y}, t_{i,z}]\Tran$ be a
tentative decision for the position of UAV $i$, $\pv_i$, and let
$\tv{\triangleq} [\tv_1\Tran,\ldots, \tv_N\Tran]\Tran$.  Then, an
estimate $\widehat{\pv}$ of the actual components $\pv$ can be obtained
by solving the problem
\begin{equation}
\label{eq:min_square_error}
\widehat{\pv} = \arg \min_{\tv} \Ec(\tv)
\end{equation} 
where
\begin{equation}
\Ec(\tv)\triangleq \sum_{i=1}^N \sum_{\substack{j=1 \\j\neq
    i}}^N\sum_{\substack{k=1 \\ k \neq
    i,j}}^{N}\left(\widetilde{\delta}_{i,j,k}-\theta_{i,j,k}(\tv)\right)^2 \label{eq:energy}
\end{equation}
is the square error between the channel observations,
$\widetilde{\delta}_{i,j,k}$, and the tentative distances
$\theta_{i,j,k}(\tv){\triangleq} |\tv_j-\tv_k| {+}
|\tv_k-\tv_i| {-} |\tv_j-\tv_i|$  according to~\eqref{eq:delta_ijk}.
The square error
in~\eqref{eq:energy} is, in general, a non-convex function,
nevertheless local or global minima can be easily found by applying a
standard GD method.

In the GD algorithm, let $\tv^{(\alpha)}$ be the tentative positions
at iteration $\alpha{\geq}0$. Then, after applying one GD step, the
tentative estimated position at iteration $\alpha{+}1$ is updated as
$\tv^{(\alpha+1)}{=}\tv^{(\alpha)}{-}\gamma^{(\alpha)}\gv^{(\alpha)}$
where $\gamma^{(\alpha)}$ denotes the step size, $\gv{\triangleq}
\frac{\partial}{\partial \tv}\Ec(\tv)$ is the gradient of $\Ec(\tv)$
and $\gv^{(\alpha)}{=}\gv |_{\tv = \tv^{(\alpha)}}$.  The step size
can be kept constant or selected at each iteration according to some
rule as, e.g., the Barzilai-Borwein method, where $\gamma^{(\alpha)}$
is computed by exploiting the trend of the most recent two
iterations~\cite{Barzilai1988}.  Let $\gv{\triangleq}
[\gv_1\Tran,\ldots,\gv_N\Tran]\Tran$. Then $\gv_h$, $h=1,\ldots,N$, is given by
\begin{equation}
\gv_h(\tv) =\frac{\partial \Ec(\tv)}{\partial \tv_h} = -2\sum_{i=1}^N
\sum_{\substack{j=1 \\j\neq i}}^N\sum_{\substack{k=1 \\ k \neq
    i,j}}^{N} w_{i,j,k}(\tv)\frac{\partial \theta_{i,j,k}(\tv) }{\partial
  \tv_h}\label{eq:g1}
\end{equation}
where $w_{i,j,k}(\tv){\triangleq}\widetilde{\delta}_{i,j,k}{-}
\theta_{i,j,k}(\tv)$.  Since $\frac{\partial |\xv|}{\partial\xv} {=}
\frac{\xv}{|\xv|}$, for any vector $\xv$, using~\eqref{eq:delta_ijk}
we have
\begin{equation}\frac{\partial \theta_{i,j,k}(\tv)}{\partial \tv_h} =  \left\{
\begin{array}{ll}
\uv_{i,k}(\tv)-\uv_{i,j}(\tv)& \mbox{if } i=h
\\ \uv_{j,k}(\tv)-\uv_{j,i}(\tv) & \mbox{if } j=h
\\ \uv_{k,i}(\tv)+\uv_{k,j}(\tv) & \mbox{if } k=h \\ \zerov &
\mbox{else}\end{array} \right.\label{eq:derivative_theta}
\end{equation}
where the versors $\uv_{n,m}$ are defined in~\eqref{eq:versors}. By
using~\eqref{eq:derivative_theta} in~\eqref{eq:g1}, we obtain
\[
\gv_h = -2 \left\{
\begin{array}{ll}
\sum_{\substack{j\neq h}}\sum_{\substack{k \neq h,j}}
w_{h,j,k}(\uv_{h,k}-\uv_{h,j}) & \mbox{if } i=h \\ \sum_{i\neq
  h}\sum_{k\neq i,h}w_{i,h,k}(\uv_{h,k}-\uv_{h,i}) &\mbox{if } j=h
\\ \sum_{i\neq h}\sum_{j\neq i,h} w_{i,j,h}(\uv_{h,i}+\uv_{h,j}) &
\mbox{if } k=h\,.
\end{array}
\right.
\]
By summing the above three contributions and renaming the summation
indices, after some algebra, we obtain
\[
\gv_h {=} 2\sum_{\substack{i=1 \\i\neq h}}^N\sum_{\substack{j=1 \\ j \neq h,i}}^N \!\!(w_{h,i,j}{+}w_{i,h,j})\!\left(\uv_{h,i}{-}\uv_{h,j}\right)-w_{i,j,h} \!\left(\uv_{h,i}{+}\uv_{h,j}\right)
\]
where, for simplicity, we omitted the dependency on $\tv$. In our
implementation, the GD algorithm starts from an initial tentative
value, $\tv^{(0)}$, and iterates until the following stopping
condition on the square error is satisfied:
\[ 
\frac{\Ec(\tv^{(\alpha+1)})-\Ec(\tv^{(\alpha)})}{\Ec(\tv^{(\alpha)})}<\epsilon\,,
\]
or a maximum number of iterations, $I_\alpha$, is reached.
The initial tentative $\tv^{(0)}$ can be (i) chosen either as a prior estimate of the position or (ii) randomly
  drawn. Case (i) is selected when TIP operates in tracking mode, or in cold star from the second TIP iteration onwards; case (ii) is selected only in cold start mode at the first TIP iteration.

\subsection{Improving convergence to global optimum}

Since the square error function $\Ec(\tv)$ is not convex, the GD
algorithm may reach a local minimum instead of the global one, thus
leading to inaccurate positioning.  To ensure convergence to the
global minimum with high probability, we propose a heuristic method
based on the following considerations.  Consider that the GD algorithm
has converged to its global minimum and that the minimizer
of~\eqref{eq:energy} is $\tv {=} \widehat{\pv}$. Such estimate of the
actual positions $\pv$ can be written as $\widehat{\pv}{=}\pv+\ev$ where
$\ev$ is the estimation error.  Hence, 
\[ 
\theta_{i,j,k}(\widehat{\pv}) = \theta_{i,j,k}(\pv) + \psi_{i,j,k}
\]
where, by definition, $\theta_{i,j,k}(\pv) {=} \delta_{i,j,k}$ and
$\psi_{i,j,k}$ is the estimation error in the RED.
Using~\eqref{eq:distance_noise}, the term in the sum
in~\eqref{eq:energy} reduces to
$\widetilde{\delta}_{i,j,k}{-}\theta_{i,j,k}(\widehat{\pv}) {=}
\eta_{i,j,k}-\psi_{i,j,k}$ where we recall that the terms
$\eta_{i,j,k}$ are the quantization errors on the channel
observations. Thus, the square error at convergence is
\begin{equation}
\Ec(\widehat{\pv})= \sum_{i=1}^N \sum_{\substack{j=1 \\j\neq
    i}}^N\sum_{\substack{k=1 \\ k \neq
    i,j}}^{N}\left(\eta_{i,j,k}-\psi_{i,j,k}\right)^2\,.\label{eq:energy2}
\end{equation}
Since we assume that the global minimum of the function $\Ec(\tv)$ has
been reached at $\tv{=}\widehat{\pv}$, any other choice for $\tv$ will
lead to a larger square error. Consider now the choice $\tv{=}\pv$,
which corresponds to $\psi_{i,j,k}{=}0$, $\forall i,j,k$. According
to~\eqref{eq:energy2}, we then have
\begin{equation} \Ec(\widehat{\pv}) \le \Ec(\pv) = \sum_{i=1}^N \sum_{\substack{j=1 \\j\neq i}}^N\sum_{\substack{k=1 \\ k
\neq i,j}}^{N}\eta_{i,j,k}^2\,.
\label{eq:bound_E}
\end{equation}
Now, given the actual position components, $\pv$, the quantization
errors, $\eta_{i,j,k}$, are deterministic quantities; however, as
already observed, the uncertainty in $\pv$ can be translated into
$\eta_{i,j,k}$ being independent random variables with zero-mean,
uniform distribution in the interval $[{-}\frac{c}{2B},\frac{c}{2B}]$
and, hence, variance $\sigma^2_\eta{=}\frac{c^2}{12B^2}$.  For
sufficiently large $N$ (that is, a sufficiently large number of
observations $\widetilde{\delta}_{i,j,k}$), the following holds with
high probability:
\begin{equation} \Ec(\widehat{\pv}) \le  N(N-1)(N-2)\frac{c^2}{12B^2} \triangleq E_b \,.
\label{eq:residual_square_error}\end{equation}
In summary, when the GD algorithm converges to the global minimum, the
residual square error is expected to be lower than $E_b$, whereas we
expect $\Ec(\widehat{\pv})$ to be much larger when it converges to a
local minimum.  Based on the above considerations, we improve the
positioning accuracy of our algorithm by comparing the square error
$\Ec(\widehat{\pv})$ at the output of the GD algorithm against the
threshold $\widetilde{E}_b {=} \beta E_b$ where $\beta{>}1$
is a constant.

Referring to Fig.~\ref{fig:cold_start}, the GD verifies the
condition $\Ec(\widehat{\pv}){>}\widetilde{E}_b$ at the $L$-th
iteration. If it is not met, the TIP is run again resetting $\ell{=}0$
and using a new randomly drawn $\widehat{\pv}^{(0)}_{\sf
  cold\,start}$.  This restart is repeated until
$\Ec(\widehat{\pv}){\le} \widetilde{E}_b$, or a maximum number of
retries has been reached. Otherwise, failure is declared.  Note that
the purpose of the constant $\beta$ is to compensate for
possible large statistical deviations of $\Ec(\pv)$ (i.e., the
r.h.s. of~\eqref{eq:bound_E}) from $E_b$.  As $\beta$ decreases,
the probability that reliable estimates are declared unreliable
increases as well as the computational complexity of the
algorithm. Conversely, as $\beta$ increases, convergence to
local minima are less likely to be detected and, in general, the
positioning accuracy decreases.

\begin{remark}
The GD algorithm proposed above allows the joint estimation of 
    the positions of all UAVs. However, we recall that only the
    positions of $\bar{N}{=}N-A$ UAVs  have to be estimated since
    the remaining $A$ are anchors whose positions are 
    perfectly known. To take this into account, the GD
    algorithm needs to be slightly modified as follows. Let $\Ac$ be the set of
    anchor nodes.  Then, for all $i{\in} \Ac$, the tentative
    decisions are set to  the
    actual positions, i.e., $\tv_i^{(0)}{=}\pv_i$ and the gradient $\gv_i^{(\alpha)}$ is set to
    $\zerov$ at every iteration step $\alpha$.  This ensures
    $\tv_i^{(\alpha)}{=}\pv_i$ for all iteration steps. 
    The GD algorithm remains unchanged since the estimation error $\ev {=} \widehat{\pv}{-}\pv$ is zero for all anchor nodes,
    while it is (in general) non-zero for the others.
\end{remark}

\subsection{Estimating the nodes' velocities\label{sec:velocities}}
Once the estimates of the UAVs' position vectors  $\widehat{\pv}$ and
of the maps $\widehat{\muvL}$ are available, the components $\vv_i$ of
the UAVs' geometric velocities $\vec{V}_i$, $i{=}1,\ldots,N$ can be
estimated through~\eqref{eq:v_ijk}.  Indeed, from~\eqref{eq:v_ijk}, we
can write
\begin{eqnarray}
\omega_{i,j,k} &=& \uv_{j,k}\Tran
\vv_j+(\uv_{k,i}-\uv_{j,k})\Tran\vv_k -\uv_{k,i}\Tran\vv_i\non &=&
\uv_{i,j,k}\Tran \vv\label{eq:omega_vector}
\end{eqnarray}
where $\vv {=} [\vv_1\Tran, \ldots, \vv_N\Tran]\Tran$.  If we collect
the terms $\omega_{i,j,k}$ and $\widetilde{\omega}_{i,j,k}$ in the
column vectors $\omegav$ and $\widetilde{\omegav}$, respectively,
according to~\eqref{eq:velocity_noise}, we can write
\begin{equation}
  \widetilde{\omegav} = \omegav +\etav = \Um\Tran\vv +\zetav\,.
  \label{eq:omegaUv}
\end{equation}
where the columns of the matrix $\Um$ are $\uv_{i,j,k}$ and the
elements of $\zetav$ are the discretization errors $\zeta_{i,j,k}$.
The matrix $\Um$ has size $3N {\times} N(N{-}1)^2$, since
$N(N{-}1)^2$ is the total number of observations and $3N$ is the total
number of components of the velocity vectors.

Since the positions
  and velocities are known for the $A$ anchors, the edge server only needs to estimate
  $3\bar{N}$ parameters, with $\bar{N}{=}N-A$. In practice, if we let $\bar{\Nc}$ be the
  set of the UAVs which are not anchors, we just need to consider the
  reduced equation
\begin{equation}
  \widetilde{\omegav} = \Umk\Tran \vvk +\zetav\,,
  \label{eq:omegaUv_reduced}
\end{equation}
instead of~\eqref{eq:omegaUv}.
In \eqref{eq:omegaUv_reduced}, $\Umk$ is a $N(N{-}1)^2 {\times} 3\bar{N}$ matrix containing only the rows of $\Um$ corresponding to
the UAVs in $\bar{\Nc}$ and, similarly, $\vvk$ has size $3\bar{N}$ and contains
only the contributions of the $\bar{N}$ UAVs in $\bar{\Nc}$.

We recall that
$\zetav$ represents the quantization errors on the observed velocities
and that, due to the rounding operation, they are deterministically
dependent upon $\omegav$ but, as explained in Sec.~\ref{sec:maps}, the
uncertainty in the value of $\omegav$ translates into $\zetav$ behaving
as i.i.d. random variables with zero mean and uniform distribution.

If no statistical information about $\vvk$ is
  available, the least square estimator
$  \widehat{\vvk} = \left(\Umk\Umk\Tran\right)^{-1}\Umk
\widetilde{\omegav}
$

could be exploited\footnote{When instead a prior distribution for $\vvk$
  is known, better estimators (e.g., the maximum a posteriori
  estimator) can be used.}.
  However, the above least square estimator cannot
  directly be applied for the following two reasons:
\begin{itemize}
\item The matrix $\Umk$ is a function of the unknown vectors
  $\pv_i$, $i\in \bar{\Nc}$, for which only estimates are available. Then,
  in practice, we need to replace $\Umk$ in the above expression  with
  the matrix $\widehat{\Umk}$ computed using the estimated
  positions $\widehat{\pv}_i$, $i\in \bar{\Nc}$, obtained  at the output of
  TIP.
\item The observations $\widetilde{\omega}_{i,j,k}$ are only available
  as the elements $v_{i,j,m}$ of the ordered lists $\Vc_{i,j}$. We
  recall that the relation between the terms
  $\widetilde{\omega}_{i,j,k}$ and $v_{i,j,m}$ is completely defined
  by the maps $\muvL$, which are unknown. Hence, in a practical
  implementation, one must replace $\widetilde{\omegav}$ with
  $\widetilde{\omegav}^\star$, obtained by reordering the lists
  $\boldsymbol{\Vc}$ using the estimated maps $\widehat{\muvL}$.
\end{itemize}
It follows that the estimate of $\vvk$ can  be obtained as
\begin{equation}
  \widehat{\vvk} =
  \left(\widehat{\Umk}\widehat{\Umk}\Tran\right)^{-1}\widehat{\Umk}
  \widetilde{\omegav}^\star\,,
   \label{eq:LS}
\end{equation}
and it is affected by three sources of errors: the quantization error
represented by the terms $\zeta_{i,j,k}$, the position error
$\pv-\widehat{\pv}$, and the error in estimating the maps $\muvL$.

\section{Turbo Iterative UAV Positioning and Tracking\label{sec:turbo}}
The TIP algorithm schematized in Fig.~\ref{fig:cold_start} combines the
methods presented in Section~\ref{sec:positions}, in a iterative solution for
estimating both positions and velocities of the UAVs. Here, we provide a detailed description of the TIP operational
modes.  In particular, first Sec.~\ref{sec:positioning}  
introduces the {\em cold
  start mode}, a version of TIP designed for the most challenging case where no prior
knowledge about positions, velocities, or trajectories of the UAVs is
available to the system. 
Then Sec.~\ref{sec:tracking}  describes the {\em tracking mode}, where UAVs
follow a trajectory and provide channel profiles to the edge server
once every $\Delta t$ seconds. At each time step, TIP exploits the
current channel profiles and the previous estimates of the positions/velocities, to infer the current positions and velocities.

\subsection{Cold start mode\label{sec:positioning}}
The pseudocode for TIP working in cold start mode is outlined in
Algorithm~\ref{alg:turbo}. TIP takes as input the set of lists $\Dcb$ and $\Vcb$, which are collected by the edge server.  We recall that
such lists contain the discretized delay-Doppler profiles of all channels connecting any pair of UAVs.  
As explained in
Sec.\,\ref{sec:system-problem}, the elements of list $\Dc_{i,j}$ are
the estimated path distances $d_{i,j,m}$, $m{=}1,\ldots,N-1$, ordered
in increasing order. The first main challenge for estimating the UAV
positions is to associate each path distance $d_{i,j,m}$ with the
identity of the UAV that generated it, i.e., to find the maps $\muvL$.
These maps cannot be obtained deterministically, rather they have to be
estimated. 

\aletext{An initial estimate, $\muvL^{(0)}$, of the maps is provided using the BP
approach described in Algorithms~\ref{alg:belief} and~\ref{alg:belief2}. At the core of the TIP algorithm there is a
sophisticated mechanism designed for iteratively refining such maps. In the following, 
we denote by $\widehat{\muvL}^{(\ell)}$ the set of estimated maps at iteration $\ell$, which  
are employed to reorder the lists $\Dcb$ and $\Vcb$ according to the rule specified in Sec.~\ref{sec:apply_maps}.
Specifically, $\widehat{\muL}^{(\ell)}_{i,j}$ is applied to $\Dc_{i,j}$ and $\Vc_{i,j}$ so as
to reorder their elements according to the UAVs' identity index. In doing so, we obtain lists
$\widehat{\Delta}^{(\ell)}_{i,j}$ and $\widehat{\Omega}^{(\ell)}_{i,j}$, respectively, where their $k$-th elements,
$\widehat{\delta}^{(\ell)}_{i,j,k}$ and $\widehat{\omega}^{(\ell)}_{i,j,k}$, are
an estimate of the distance $\delta_{i,j,k}$ and of the velocity
$\omega_{i,j,k}$, respectively. If the map is perfectly estimated, i.e.,
$\widehat{\muL}^{(\ell)}_{i,j}{=}\muL_{i,j}$, then
$\widehat{\delta}^{(\ell)}_{i,j,k}{=}\widetilde{\delta}_{i,j,k}$ and
$\widehat{\omega}^{(\ell)}_{i,j,k}{=}\widetilde{\omega}_{i,j,k}$.

The set of reordered lists,
$\widehat{\Deltam}^{(\ell)}{=}\{\widehat{\Delta}^{(\ell)}_{i,j}|
i,j{=}1\ldots,N,i\neq j\}$, is then fed to the GD algorithm described in
Sec.~\ref{sec:gradient_descent}. The GD algorithm is initialized with estimate $\widehat{\pv}^{(\ell)}$ at iteration $\ell>0$, and with a random vector $\widehat{\pv}_{\sf init}$ at iteration $\ell=0$.}
The GD algorithm then outputs the estimated positions, $\widehat{\pv}^{(\ell+1)}$, which are
employed to 
(i) provide an estimate of the velocities, $\widehat{\vv}^{(\ell{+}1)}$,
using the reordered set of lists $\widehat{\Omegam}^{(\ell)}{=} \{\widehat{\Omega}^{(\ell)}_{i,j}| i,j{=}1\ldots,N, i {\neq} j \}$
and the procedure described in Sec.~\ref{sec:velocities} and
(ii) update the estimate of the maps.

This latter task, performed by the block labeled ``Compute maps'' in
Fig.~\ref{fig:cold_start}, follows the procedure
outlined below:
\begin{enumerate}
\item from the estimated positions $\widehat{\pv}^{(\ell)}$ and for each $i{\neq}j$, we compute the distances $\delta^\star_{i,j,k}$  using~\eqref{eq:delta_ijk}, as
  \begin{equation}  \delta^\star_{i,j,k} {=}|\widehat{\pv}^{(\ell)}_j-\widehat{\pv}^{(\ell)}_k| {+} |\widehat{\pv}^{(\ell)}_k{-}\widehat{\pv}^{(\ell)}_i|{-}|\widehat{\pv}^{(\ell)}_j {-} \widehat{\pv}^{(\ell)}_i|\,,\label{eq:delta_ijk_star}
  \end{equation}
  for $k{\in} \{1,\ldots, N\}{\setminus}\{i\}$, and we arrange them in the list
  $\Delta^\star_{i,j}$;
\item the elements of $\Delta^\star_{i,j}$ are then ordered in
  ascending order to form the list $\widetilde{\Delta}^\star_{i,j}$;
\item finally, the new estimated map $\widehat{\muL}^{(\ell)}_{i,j}$ is defined as
  the rule that transforms $\Delta^\star_{i,j}$ into
  $\widetilde{\Delta}^\star_{i,j}$.
\end{enumerate} 
The new maps are then employed to provide a better reordering of the
lists $\Dcb$ and $\Vcb$. The algorithm proceeds iteratively until a
desired number of iterations, $L$, has been performed. 

\begin{algorithm} 
\caption{TIP algorithm: Cold start mode}\label{alg:turbo}
\begin{algorithmic} 
  \REQUIRE $\Dcb$, $\Vcb$, $L$, $\widehat{\muvL}^{(0)} {\gets} {\rm EstimateMaps}(\Dcb)$, $\widehat{\pv}_{\sf init}{\gets} {\rm rand}$
  \FOR{ $\ell \gets 0$ to $L-1$}
     \IF{$\ell=0$}
         \STATE $\widehat{\pv}^{(0)} \gets \widehat{\pv}_{\sf init}$
     \ELSE
\FOR{$i,j=1,\ldots,N$, $j\neq i$}
  \STATE Create the lists $\Delta^\star_{i,j}$ using $\widehat{\pv}^{(\ell)}$
and~\eqref{eq:delta_ijk_star}
  \STATE Sort each list $\Delta^\star_{i,j}$ in ascending order to obtain
  $\widetilde{\Delta}^\star_{i,j}$
  \STATE Compute the map $\widehat{\muL}_{i,j}^{(\ell)}$ that yield
  $\widetilde{\Delta}_{i,j}^\star$ from $\Delta_{i,j}^\star$
  \ENDFOR
     \ENDIF
\STATE Apply the maps $\widehat{\muvL}^{(\ell)}$ to
$\boldsymbol{\Dc}$ to obtain $\widehat{\Deltam}^{(\ell)}$
\STATE Apply
the maps $\widehat{\muvL}^{(\ell)}$ to $\boldsymbol{\Vc}$ to obtain
$\widehat{\Omegam}^{(\ell)}$ \STATE $\widehat{\pv}^{(\ell+1)}
\gets$ {\rm GradientDescent}$(\widehat{\Deltam}^{(\ell)},\widehat{\pv}^{(\ell)})$
\STATE Compute $\widehat{\bar{\vv}}^{(\ell+1)}$ using~\eqref{eq:LS}

\STATE $\ell\gets \ell+1$ \ENDFOR \RETURN
$\widehat{\pv}^{(L)}$
\end{algorithmic} 
\end{algorithm}

\subsection{Tracking mode\label{sec:tracking}}
We now consider the case where each UAV follows a trajectory
in space that can be described by the position vectors
$\vec{P}_i(t)$ and velocity vectors $\vec{V}_i(t){=}\frac{\dd}{\dd
t}\vec{P}_i(t)$, which are functions of time $t$. While moving along
the trajectory, UAVs periodically send to the edge server (say every
$\Delta t$ seconds) the channel profiles obtained through the 
OTFS channel estimation,
{\color{black} which occurs within a time frame of duration.
In such a short time we do not expect position and velocity of the UAVs to change significantly. }
  
Then, at time $t$, TIP takes as input the set of lists 
$\Dcb(t)$ and $\Vcb(t)$ and provides an estimate of the positions
$\widehat{\pv}(t)$ and velocities $\widehat{\vv}(t)$. To ease this task, 
the algorithm also exploits an estimate of the UAV positions
previously obtained at time $t-\Delta t$.

In this scenario,  TIP  is connected to the tracking module depicted in Fig.~\ref{fig:tracking}. Also, since a prior estimate
of the UAV positions, $\pv_{\sf init}(t)$, is already available, the block named ``Belief propagation'' is not activated. The reason for this choice is that BP is able to infer
reliable estimates of the maps when no prior information on the UAVs'
positions are available at a price of computational complexity
$O(N^8)$. However, in tracking mode, the system has prior estimates of the UAVs' positions, which makes the use of BP
unnecessary.  Apart from this aspect, the tracking module works
as described in Sec.~\ref{sec:positioning}.

Summarizing, in tracking mode, TIP works as follows. At time $t$, TIP
\begin{itemize}
  \item gets as input the channel profiles $\Dcb(t)$ and $\Vcb(t)$, as
    well as the prior estimate of the positions $\pv_{\sf init}(t)$
    obtained at time $t{-}\Delta t$;
  \item provides a current estimate of the positions $\widehat{\pv}(t)$ and
    of the velocities $\widehat{\vv}(t)$;
  \item the tracking module provides a forecast for the UAV positions at time
    $t{+}\Delta t$, computed as
\[  \pv_{\sf init}(t+\Delta t) = \widehat{\pv}(t) +\Delta t\widehat{\vv}(t) \,.\] 
\end{itemize}

\subsection{Genie aided TIP\label{sec:GA}}
As a benchmark for the performance of our proposed  TIP, 
we consider a genie-aided (GA) version of TIP which has perfect knowledge of the maps $\muvL$.
The structure of the GA is a simplification of the scheme in Fig.~\ref{fig:cold_start}
where $L{=}0$ (no TIP iterations are performed) and the block ``Apply maps'' takes as input the actual maps, i.e.,
$\widehat{\muvL}^{(0)}{=}\muvL$. Consequently, the block ``Belief
propagation'' is not required and the TIP block  ``Compute maps'' is unnecessary.
\ematextz{As shown in Sec.\,\ref{sec:results}, the GA version of  TIP will be used to provide the lower bound on the error for positions and velocities, since  error-free maps are used.}

{\color{black} \subsection{Overall solution complexity}
We now discuss the complexity of our proposed solution. As mentioned in Sec.~\ref{sec:maps}, the  BP algorithm designed to associate paths with UAV identities has complexity $O(N^8)$. However, this does not prevent its application to large swarms of UAVs. Large UAV swarms can be handled using a frequency division approach. In practice, a swarm composed of $N$ UAVs can be first partitioned into smaller groups of $M$ UAVs each, and, to avoid interference, a different frequency band can be assigned to each group for communication. By doing so, the complexity of the BP algorithm applied to each
group is $O(M^8)$. Then, for constant $M$, the overall complexity of the BP algorithm is just $O(N)$. The assignment of UAVs to groups can later be modified to increase diversity and improve localization performance in the tracking mode.

For the complexity of the TIP algorithm, we refer to Fig.~\ref{fig:cold_start}.
\begin{itemize}
\item The position estimation block implements the gradient descent algorithm, which iteratively updates the energy $\Ec(\pv)$ in~\eqref{eq:energy} and the tentative positions $\tv$. The complexity of computing $\Ec(\pv)$ and all gradients $\gv_h$ is $O(N^3)$ as three summations from 1 to $N$ are involved in their expressions. The gradient descent algorithm stops after reaching convergence. This requires a variable number of iterations, $I_{\alpha}$, depending on the values $\widetilde{\delta}_{i,j,k}$. However, $I_{\alpha}$ does not depend on $N$, and we can limit it to a maximum value.

\item The complexity of estimating the velocities is due to the computation of~\eqref{eq:LS}, which requires the inversion of a $3\bar{N}\times 3\bar{N}$ matrix and, thus has complexity $O(N^3)$.

\item The block ``Apply maps" has also complexity $O(N^3)$, since it involves the permutation of $N(N-1)$ lists of $N-1$ elements.

\item Finally, the ``Compute Maps" block follows the procedure outlined in Sec.~\ref{sec:positioning}, which evaluates the estimated distances~\eqref{eq:delta_ijk_star} for all $i,j,k\in 1,\ldots,N$ and then sorts the lists $\Delta^\star_{i,j}$. Again, the complexity of this block is $O(N^3)$. 
\end{itemize}

In conclusion, the overall complexity of TIP is  $O(N^3)$. We demonstrate the convergence of BP and TIP in the numerical examples in Sec.~\ref{sec:results}.
}

\section{Cram\'er Rao lower bounds\label{sec:CRLB}}
To provide a benchmark for the performance of the proposed
  TIP algorithm, we now derive the joint Cram\`er Rao lower bound
  (CRLB) to the variance of the UAVs' position and velocity estimates.
  To this end, we assume that the set of maps $\muvL$ is
  perfectly known, so that it is possible to correctly associate the
  elements of the lists $\Dc_{i,j}$ and $\Vc_{i,j}$ with the
  corresponding UAVs' identities.

We recall that $A$ UAV out of $N$ are anchors, i.e., their positions
and velocities are perfectly known. Therefore, the joint
CRLB should only refer to the $\bar{N}{=}N{-}A$ UAVs which are not anchors.  Let
$\bar{\Nc}$ be the set of such UAVs and define $\pvk$ and $\vvk$ as the 
$3\bar{N}$-size vectors obtained by stacking the positions and velocities, $\pv_i$
and $\vv_i$, respectively, $\forall i{\in} \bar{\Nc}$.  In the CRLB
terminology, the position and velocities $\bar{\pv}$ and $\bar{\vv}$
represent the parameters to be estimated, while the set of
distances $\Dcb$ and velocities $\Vcb$ are the observations. Hence, the total
number of parameters to estimate is $6\bar{N}$ i.e., 3 position
components and 3 velocity components for each UAV in $\bar{\Nc}$.  

Let us first consider the UAV's positions and velocities
vectors as random variables, whose components are i.i.d. and have
density $f_p(p)$ and $f_v(v)$, respectively. In other words, the joint density of the elements of $\pvk$ is
$f_{\pvk}(\pvk){=}\prod_{\ell=1}^{3\bar{N}}f_p(\bar{p}_{\ell})$ and, similarly,
$f_{\vvk}(\vvk){=}\prod_{\ell=1}^{3\bar{N}}f_v(\bar{v}_{\ell})$ where
$\bar{p}_{\ell}$ and $\bar{v}_{\ell}$ are the $\ell$-th elements of $\pvk$
and $\vvk$, respectively.

As already discussed in Sec.~\ref{sec:maps}, although the
discretization errors $\eta_{i,j,k}$ and $\zeta_{i,j,k}$ appearing
in~\eqref{eq:distance_noise} and~\eqref{eq:velocity_noise} are
functions of (hence, correlated with) $\delta_{i,j,k}$ and $v_{i,j,k}$ (resp.), the uncertainty in the UAVs' positions and velocities
can be translated into $\eta_{i,j,k}$ and $\zeta_{i,j,k}$ being
i.i.d. random variables, with distributions $f_{\eta}(\eta)$ and
$f_{\zeta}(\zeta)$ uniform in the ranges $[-\frac{c}{2B},
  \frac{c}{2B}]$ and $[-\frac{c}{2f_cT_f},\frac{c}{2f_c T_f}]$,
respectively.  Such an assumption allows us to define the joint density of observations and parameters as
$f_{\Dcb,\Vcb,\pvk,\vvk}(\Dcb,\Vcb,\pvk,\vvk)$.

It follows that the joint CRLB on the $\ell$-th parameter to be estimated is given by $[\Fm^{-1}]_{\ell,\ell}$ where $\Fm$ is the
$6\bar{N}\times 6\bar{N}$ Fisher information matrix (FIM) defined as $\Fm {=}
\EE_{\Dcb,\Vcb, \pvk,\vvk}\left[\yv\yv\Tran\right]$ where
\begin{equation}  \yv = \left[\begin{array}{c} \frac{\partial}{\partial \pvk}\log f_{\Dcb,\Vcb,\pvk,\vvk}(\Dcb,\Vcb, \pvk,\vvk)\\ \frac{\partial}{\partial \vvk}\log f_{\Dcb,\Vcb,\pvk,\vvk}(\Dcb,\Vcb, \pvk,\vvk)  \end{array}\right]\,.\label{eq:y} 
\end{equation}
The derivation of the expression of the FIM is quite tedious,
and we only provide the final result, 
 given by: 
\[\Fm = 
\left[\begin{array}{cc} C_{\eta} \Dm_{p,p}+ C_{\zeta} \Vm_{p,p}+C_p
    \Id & C_{\zeta} \Vm_{p,v} \\ C_{\zeta} \Vm_{p,v}\Tran &
    C_{\zeta}\Vm_{v,v}+C_v \Id \end{array}\right]
\]
where 
\begin{eqnarray*}
\Dm_{p,p} &=& \sum_{i,j\neq i, k\neq \{i,j\}}\EE_{\pvk}\left[\frac{\partial \delta_{i,j,k}}{\partial \pvk}\frac{\partial \delta_{i,j,k}}{\partial \pvk\Tran}\right]\,, \nonumber\\
\Vm_{p,p} &=& \sum_{i,j\neq i, k\neq i}\EE_{\pvk,\vvk}\left[\frac{\partial v_{i,j,k}}{\partial \pvk}\frac{\partial v_{i,j,k}}{\partial \pvk\Tran}\right]\,,\non
\Vm_{p,v} &=& \sum_{i,j\neq i, k\neq i}\EE_{\pvk,\vvk}\left[\frac{\partial v_{i,j,k}}{\partial \pvk}\frac{\partial v_{i,j,k}}{\partial \vvk\Tran}\right]\,,\non
\Vm_{v,v} &=& \sum_{i,j\neq i, k\neq i}\EE_{\pvk,\vvk}\left[\frac{\partial v_{i,j,k}}{\partial \vvk}\frac{\partial v_{i,j,k}}{\partial \vvk\Tran}\right]\,, \non
\end{eqnarray*}
\begin{equation}
  C_x = \int_{-\infty}^{\infty}\frac{(f_x'(z))^2}{f_x(z)}\dd z\,,\label{eq:Cx}
\end{equation} 
with $x{\in} \{p,v,\eta,\zeta\}$.  We point out that when $f_x(x)$ is a
uniform distribution, the coefficient $C_x$ cannot be computed, due
to the discontinuities in its expression. It follows that the joint
CRLB cannot be computed under the hypothesis of uniformly distributed
$\eta_{i,j,k}$ and $\zeta_{i,j,k}$. To circumvent this problem, and
only for   the CRLB evaluation, we  assume
$\eta_{i,j,k}{\sim} \Nc(0,\sigma^2_\eta)$ and $\zeta_{i,j,k}{\sim}
\Nc(0,\sigma^2_\zeta)$ where $\sigma_\eta {=} \frac{c}{\sqrt{12}B}$
and $\sigma_\zeta {=}\frac{c}{\sqrt{12}f_c T_f}$ so that they have the
same variance as their uniformly distributed counterparts. For a
Gaussian distribution, the coefficient $C_x$ is given by $C_x{=}1/\sigma^2_x$.

The average joint {\em per-component} CRLB  on the estimate of the
position and velocity vectors are thus given by
\begin{eqnarray}  
\overline{\rm CRLB}^{\rm joint}_p &=& \frac{1}{3\bar{N}}\sum_{\ell=1}^{3\bar{N}}\left[\Fm^{-1}\right]_{\ell,\ell} \label{eq:jointCRLBp}\\
\overline{\rm CRLB}^{\rm joint}_v &=& \frac{1}{3\bar{N}}\sum_{\ell=3\bar{N}+1}^{6\bar{N}}\left[\Fm^{-1}\right]_{\ell,\ell}\,. \label{eq:jointCRLBv}
\end{eqnarray} 

\section{Numerical Results\label{sec:results}}

   We now measure the
    performance of TIP in some test scenarios. As a performance metric, we consider the
    root mean square error (RMSE) on the position and velocity
    estimate per UAV and per dimension $x,y,z$. By recalling that $\bar{\Nc}$
    is the set of non-anchor UAVs, the RMSE on the position estimates
    is computed by averaging the results obtained from $R$ runs of the
    TIP algorithm, as
\begin{equation}  {\rm RMSE}_p = \sqrt{\frac{1}{3\bar{N}R}\sum_{i\in \bar{\Nc}}\sum_{r=1}^R\left|\widehat{\pv}_{i,r} -\pv_{i,r} \right|^2}\label{eq:RMSEp}
  \end{equation}
where $\pv_{i,r}$ is the $r$-th realization of the positions $\pv_i$
and $\widehat{\pv}_{i,r}$ is the corresponding estimate.  Likewise,
the RMSE on the velocity estimates is computed as
\begin{equation} {\rm RMSE}_v = \sqrt{\frac{1}{3\bar{N}R}\sum_{i\in \bar{\Nc}}\sum_{r=1}^R\left|\widehat{\vv}_{i,r} -\vv_{i,r} \right|^2}\,.\label{eq:RMSEv}
  \end{equation}

  As a test scenario, we consider a swarm composed of $\bar{N}{=}4$ UAVs
  moving in a flight area and  $A{=}4$ anchors with fixed positions (their velocity is zero). The  $N{=}\bar{N}{+}A$ UAVs plus anchors communicate among each other using
  signals with bandwidth $B$ and central frequency $f_c{=}5$\,GHz. 
  Specifically, we assume that anchors have
  identities $i{=}1,2,3,4$, and their positions, measured in meters,
  are $\vec{P}_1{=}\vec{0}$, $\vec{P}_2{=}1000\,\vec{e}_x$,
  $\vec{P}_3{=}1000\,\vec{e}_y$, and $\vec{P}_4{=}1000\,\vec{e}_z$. 
 
  In the gradient descent algorithm, we set the stopping threshold to
  $\epsilon{=}10^{-4}$ and the maximum number of iterations to 
  $I_\alpha=100$. For the gradient step $\gamma$, we employ the
  adaptive Barzilai-Borwein method~\cite{Barzilai1988}, which
  ensures a faster convergence. Also, we set the threshold $\widetilde{E}_b {=} 2 E_b$ in~\eqref{eq:residual_square_error}.

\subsection{Cold start mode}
We start by showing the performance of TIP in cold start mode.  The results are obtained by averaging the output of TIP over
$R{=}100$ runs.  At each run, the components of the UAV's position
vectors were randomly drawn from  ${\cal N}(\mu,\sigma^2)$ with 
$\mu{=}500$\,m and  $\sigma{=}\frac{1,000}{\sqrt{12}}\approx
289$\,m.\footnote{These are the same $\mu$ and $\sigma$ of a uniform distribution of UAVs in a cube of side 1,000\,m
with vertices coinciding with the four anchor nodes.}
Also, at
each run, TIP is initialized by drawing each component of the tentative
positions $\widehat{\pv}^{(0)}_{\sf cold\,start}$ from the same Gaussian distribution.  Likewise, at each run, the
components of the UAV velocity vectors were randomly drawn from ${\cal N}(0,\sigma^2)$ with $\sigma {=} 10$\,m/s.

  \begin{figure}[t]
  \centering
  \includegraphics[width=0.50\columnwidth]{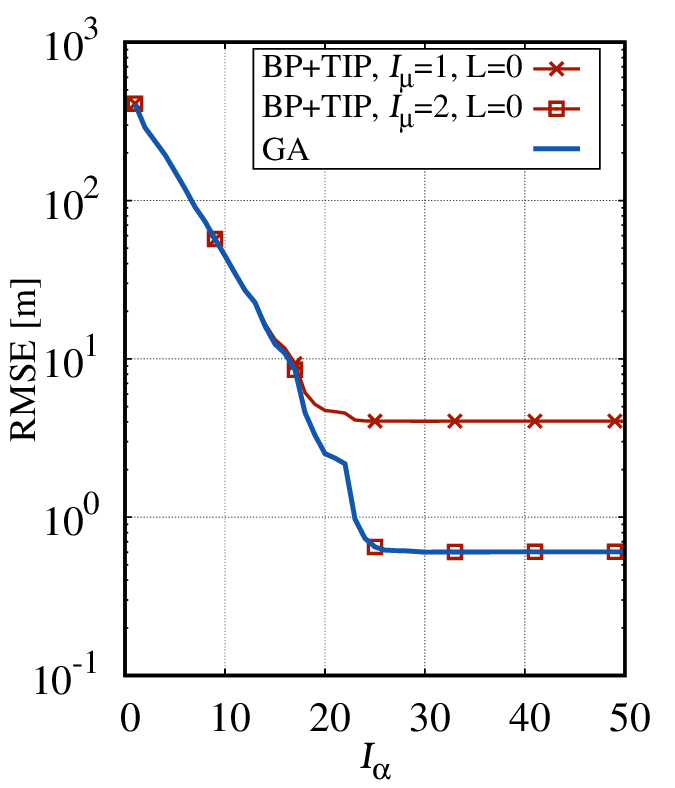}\includegraphics[width=0.50\columnwidth]{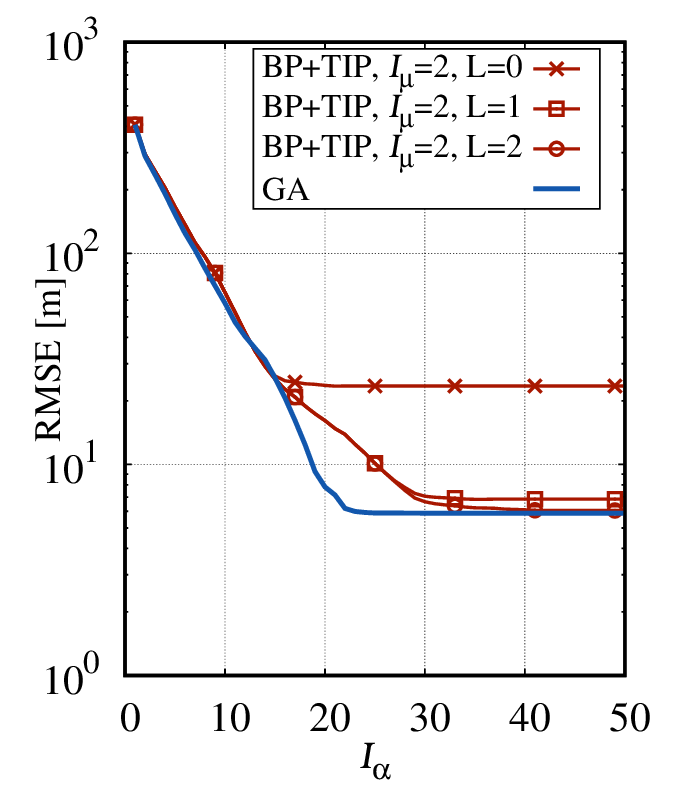}\vspace{-3mm}
\caption{RMSE as a function of the number of gradient descent
  iterations ($I_\alpha$). Left: $L{=}0$ and $B{=}30$\,MHz
  corresponding to $c\Delta\tau{=}10$\,m. Right: $I_{\mu}{=}2$ and
  $B{=}3$\,MHz corresponding to $c\Delta\tau{=}100$\,m.}
\label{fig:RMSE1}
\end{figure}

Fig.~\ref{fig:RMSE1}(left) shows the ${\rm RMSE}_p$
in~\eqref{eq:RMSEp} achieved by TIP, plotted versus the number of
iterations of the gradient descend algorithm, $I_{\alpha}$, for
$L{=}0$ (no TIP iterations) and signal bandwidth $B{=}30$\,MHz,
corresponding to a discretization step $c/B{=}10$\,m.  One can notice
that $I_{\mu}{=}2$ iterations of the BP algorithm provide the same
performance as the genie-aided (GA) algorithm, which has perfect
knowledge of the maps $\muvL$. We recall that GA is used here as
a benchmark since it represents a lower bound for the TIP ${\rm
  RMSE}_p$.  Fig.~\ref{fig:RMSE1}(left) underlines that, despite the 10\,m
discretization step in the measurements, the system achieves an ${\rm
  RMSE}_p$ of about 1\,m after $I_{\alpha}{=}30$ iterations, which
means that BP provides very reliable estimates $\widehat{\muvL}$.
However, for a smaller bandwidth, $B{=}3$\,MHz, such estimates have
much lower reliability: some TIP iterations are then required to
improve the ${\rm RMSE}_p$, as in the example shown in
Fig.~\ref{fig:RMSE1}(right).  Here the discretization step is
$c/B{=}100$\,m and the number of BP iterations is $I_{\mu}{=}2$.  For
$L{=}0$ (no TIP iterations), the resulting ${\rm RMSE}_p$ is quite
large, about 22\,m, while a single TIP iteration lowers it at about
7\,m.  Importantly, with a further TIP iteration ($L{=}2$) and
$I_\alpha{>}40$ iterations, the system reaches the GA performance.

\begin{figure} 
  \centering
  \includegraphics[width=0.49\columnwidth]{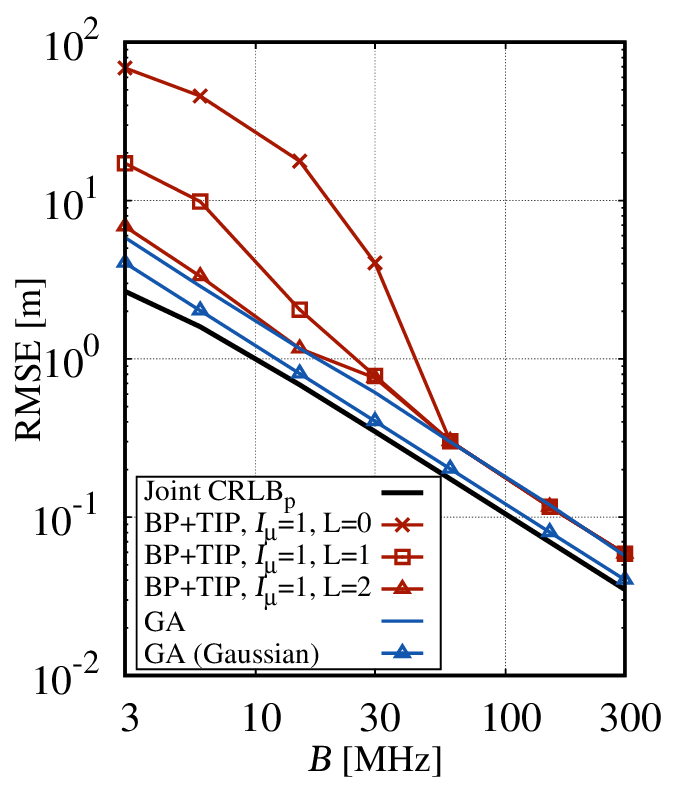}
  \includegraphics[width=0.49\columnwidth]{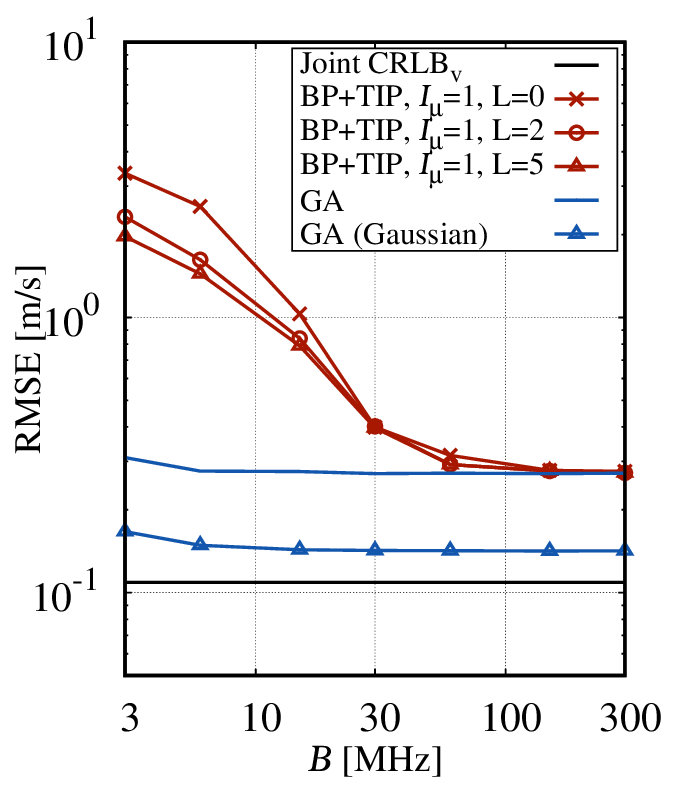}\vspace{-3mm}
\caption{RMSE on position estimates (left) and velocity estimates (right), as a function of the signal bandwidth, for varying $L$, $T_f{=}20$\,ms, and $I_{\mu}{=}1$. Average over $R{=}100$ random network scenarios.\label{fig:RMSE2}}
\end{figure}

Fig.~\ref{fig:RMSE2}(left) presents the ${\rm RMSE}_p$
in~\eqref{eq:RMSEp} versus the signal bandwidth, $B$, as the number of
TIP iterations, $L$, varies, for $I_\mu{=}1$. For the sake of
comparison, the figure also shows the performance of the GA algorithm,
and the joint CRLB$_p$ computed according to~\eqref{eq:jointCRLBp}.  We
observe that, as $B$ increases, the discretization step decreases and
the system provides, in general, more accurate positioning.  For
$L{=}2$, the system performance reaches the GA for all considered
values of $B$ in the range 3--300\,MHz.
For the same 
system parameters and for frame time $T_f{=}20$\,ms,
Fig.~\ref{fig:RMSE2}(right) shows the ${\rm RMSE}_v$
in~\eqref{eq:RMSEv} on the estimate of the UAV velocities, measured in
m/s, plotted versus the signal bandwidth.  For such a value of $T_f$,
the discretization step on the velocities is $c/f_cT_f{=}3$\,m/s.
\aletext{In this setting, the best performance is  obtained using
  $L{=}5$ TIP iterations but no further
  improvement is obtained by increasing $L$.  However, for
  $B{=}300$\,MHz, the achieved ${\rm RMSE}_v{\approx}0.27$\,m/s, which
  is well below the discretization step.

  \noindent {\em Remark 1} The gap between GA and the CRLB in both plots of 
  Fig.~\ref{fig:RMSE2} is mainly due to the fact that the quantization
  errors $\eta_{i,j,k}$ and $\zeta_{i,j,k}$ introduced
  in~\eqref{eq:distance_noise} and ~\eqref{eq:velocity_noise},
  respectively, are due to the rounding operation and depend
  upon the actual value of $\delta_{i,j,k}$ and $\omega_{i,j,k}$.
  Instead, the CRLB is computed under the assumption of Gaussian
  errors $\eta_{i,j,k}$ and $\zeta_{i,j,k}$, which are independent of
  $\delta_{i,j,k}$ and $\omega_{i,j,k}$, respectively.
 {\color{black} Indeed, by feeding the TIP algorithm with channel measurements corrupted by Gaussian noise instead of quantization noise, the performance gap is very much reduced, as shown by the curves in Figures~\ref{fig:RMSE2} (left and right) labeled as "GA (Gaussian)". The remaining small gap
is due to the non-optimality of the position/velocity estimation, which in our solution is implemented as two separate blocks, while in an optimal design position and velocity estimation could be implemented jointly at
the cost of much higher complexity.}
  
 \noindent {\em Remark 2} We have observed that the Fisher information matrix (FIM) is dominantly block diagonal for the considered values of the system
  parameters, and this explains why CRLB$_v$ in Fig.~\ref{fig:RMSE2}(right) is almost independent of
  the signal bandwidth.}

\begin{figure}[t]
  \centering
  \includegraphics[width=0.49\columnwidth]{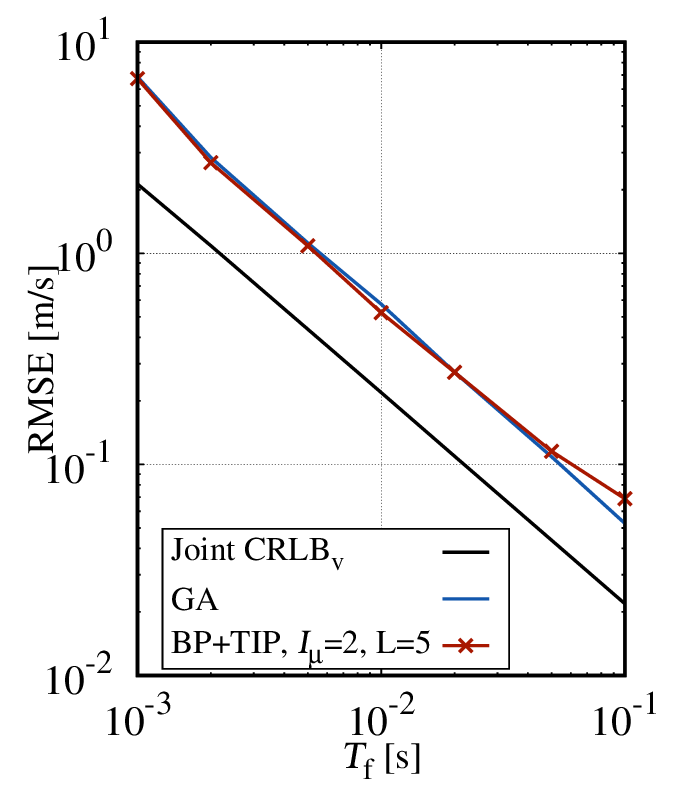}\includegraphics[width=0.49\columnwidth]{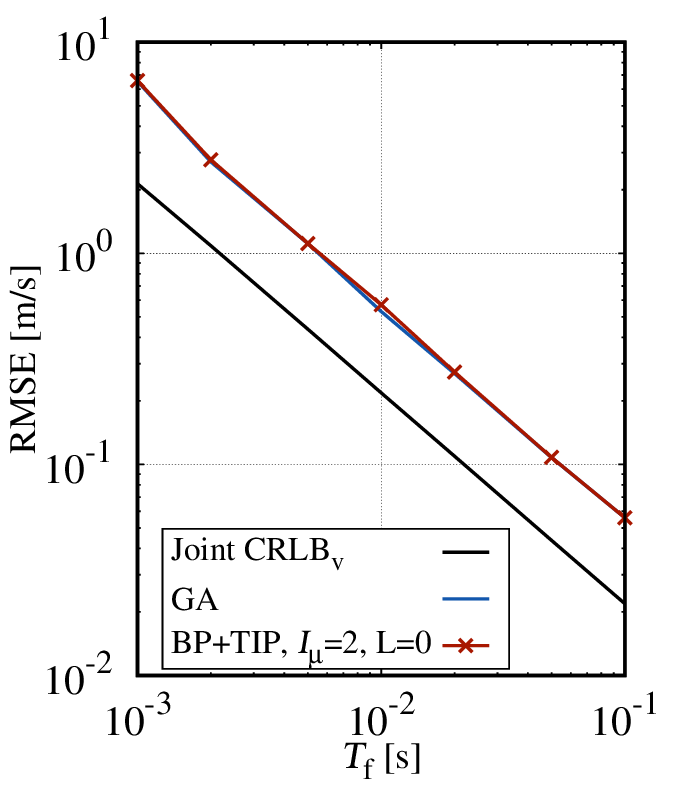}\vspace{-4mm}
\caption{RMSE on velocity estimation as a function of the frame duration,  
  for $B=10$\,MHz (left) and $B=30$\,MHz (right).}\vspace{-4mm}
\label{fig:RMSE2_velocity}
\end{figure} 

\ematextz{Fig.~\ref{fig:RMSE2_velocity}  illustrates the
performance of BP+TIP, of the GA algorithm, and of the joint CRLB computed as
in~\eqref{eq:jointCRLBv}, for $B=10$\,MHz (left) and $B=30$\,MHz (right).
The plots show that the BP+TIP
algorithm performs very close to the GA for a range of $T_f$ for any $B{\geq} 10$\,MHz.} \aletext{Similar to Fig.~\ref{fig:RMSE2}, the gap between GA and joint CRLB is due to the fact that quantization noise is not Gaussian and is correlated to the data.}
 
\begin{figure}[t]
  \centering
  \includegraphics[width=0.50\columnwidth]{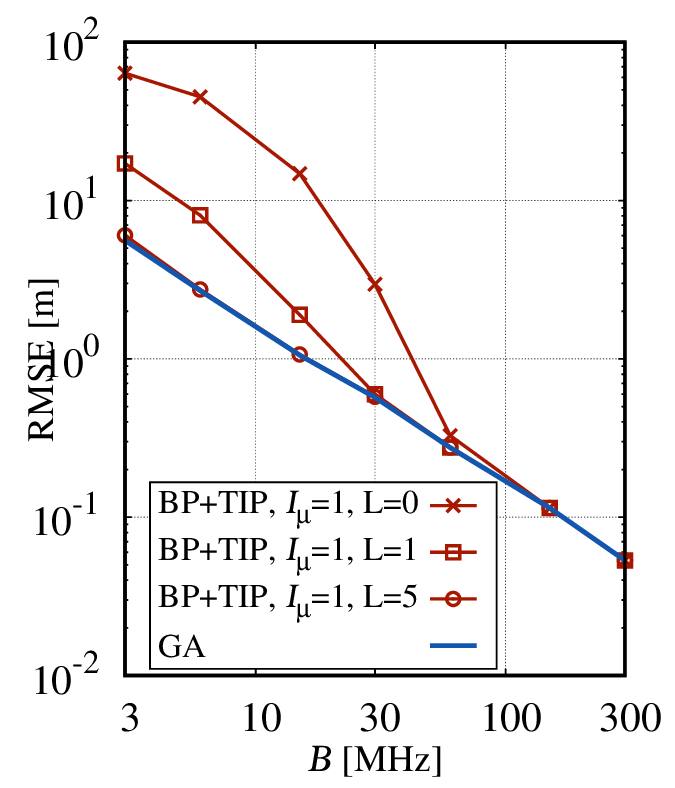}\includegraphics[width=0.50\columnwidth]{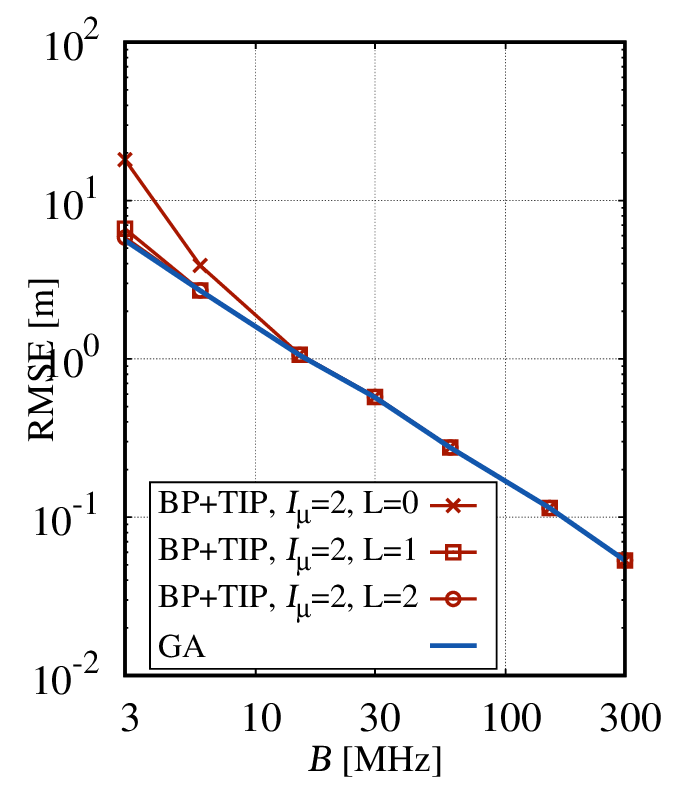}\vspace{-4mm}
\caption{RMSE on position estimation as a function of the signal bandwidth as $L$
varies, for $I_{\mu}{=}1$ (left) and $I_{\mu}{=}2$ (right). Average
over UAV positions along real-world trajectories from~\cite{UAVdataset}.}\vspace{-4mm}
\label{fig:RMSE3}
\end{figure}

 Fig.~\ref{fig:RMSE3} presents the RMSE on the position estimates,
 computed by averaging over the set of real UAV positions available
 in~\cite{UAVdataset}, whose coordinates have been linearly scaled to fit in a cube of side 1,000\,m centered at
 $[500,500,500]$\,m. All other parameters are the same as in
 Fig.~\ref{fig:RMSE2}, and  all curves are also very similar,  even though the UAV locations are not
 random. This demonstrates that our theoretical analysis does apply to
 real-world scenarios.

\subsection{Tracking mode}
To assess the performance of TIP in tracking
  mode, we consider a test scenario where the UAVs follow trajectories
  that are deterministic functions of time.  As an example of such
  trajectories, we choose the 3D Lissajous curve~\cite{Lissajous}, 
  which exhibits a simple and easy-to-implement parametric expression. The
  position vector of UAV $i$ following a Lissajous trajectory is
$\vec{P}_i(t) = p_{i,x}(t)\vec{e}_x + p_{i,y}(t)\vec{e}_y +p_{i,z}(t)\vec{e}_z $
whose components are given by:
$p_{i,s}(t) {=} a_{i,s}\sin(b_{i,s}t+\phi_{i,s})$
where $a_{i,s}$, $b_{i,s}$, and $\phi_{i,s}$ are parameters and
$s {\in} \{x,y,z\}$.
As a consequence, the instantaneous velocity vector of the $i$-th UAV is given by
$\vec{V}_i(t) {=} v_{i,x}(t)\vec{e}_x + v_{i,y}(t)\vec{e}_y
+v_{i,z}(t)\vec{e}_z$ with components 
\begin{equation}
v_{i,s}(t) = \frac{\dd p_{i,s}(t)}{\dd t} =
a_{i,s}b_{i,s}\cos(b_{i,s}t+\phi_{i,s})
\label{eq:lissajou_v}
\end{equation}
for $s {\in} \{x,y,z\}$.  

\begin{figure}[t]
  \centering
 \resizebox{0.5\columnwidth}{!}{\input{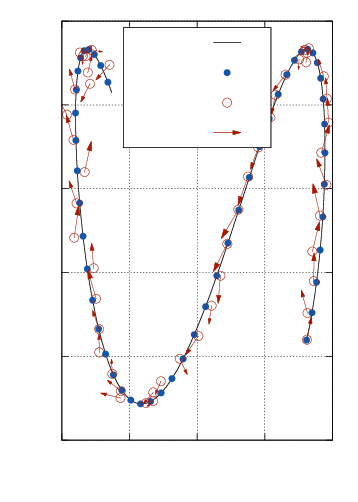}}\resizebox{0.5\columnwidth}{!}{\input{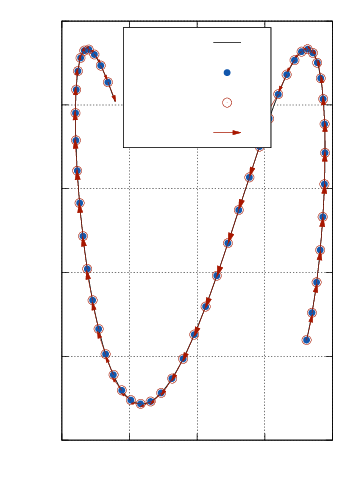}}\vspace{-4mm}
\caption{2D projection of a portion of a UAV Lissajou trajectory, for
  $N{=}8$, $A{=}4$, $B{=}3$\,MHz (left) and $B{=}300$\,MHz (right),
  $T_f{=}20$\,ms, and $\Delta t{=}1$\,s. The solid black line represents the trajectory,
  while blue circles indicate the actual UAV positions, $\vec{P}_i(n\Delta t)$. The red circles
  denote the estimated positions $\widehat{\vec{P}}_i(n\Delta t)$, and the arrows represent the estimated velocities 
  $\widehat{\vec{V}}_i(t)$.}
\label{fig:quiver}
\end{figure}

Fig.~\ref{fig:quiver}
depicts a
2D projection of a portion of the same 3D Lissajous trajectory.  This
is an instance of a random trajectory obtained by independently drawing
the parameters $a_{i,s}$ from the uniform distribution $\Uc[0,1]$,
$b_{i,s}$ from $\Uc[0,0.2]$, and $\phi_{i,\ell}$ from $\Uc[0,2\pi]$. 
The signal bandwidth is set to $B{=}3$\,MHz and $B{=}300$\,MHz, in Fig.~\ref{fig:quiver}(left) and Fig.~\ref{fig:quiver}(right), respectively. 
In both figures, $T_f{=}20$\,ms, and the update time is
$\Delta t{=}1$\,s.  In the plots, the UAV trajectory is represented by the
solid black line, while blue circles refer to the UAV position,
$\vec{P}_i(t)$ at $t{=}n \Delta t$, $n{=}0,1,{\ldots},50$.  The red circles
denote the positions $\widehat{\vec{P}}_i(n\Delta t)$ estimated by TIP for $L{=}5$ and
the arrows represent the estimated velocity vectors
$\widehat{\vec{V}}_i(n \Delta t)$. The actual velocity vectors
$\vec{V}_i(n \Delta t)$ (not shown) are tangent to the black line at
$t{=}n \Delta t$. The trajectory is traveled by the UAV from right to
left and the blue circles mark 1\,s-time intervals. For $B{=}3$\,MHz, 
the positions are  fairly well estimated (i.e., the red circles are
always close to the blue ones). However, errors in estimating the
maps sometimes result in a poor estimate of the velocity
vectors, in both direction and magnitude. By increasing the bandwidth
to 300\,MHz, the map estimates become very reliable: position estimates are excellent
and the velocity vectors perfectly follow the tangent to the trajectory.

\section{Conclusions\label{sec:concl}}
We proposed an iterative algorithm, named TIP, for localizing and tracking the position of UAVs communicating
with each other, by using RED measurements obtained through
OTFS-modulated signals. TIP exploits belief propagation and
gradient descent optimization, to achieve precise positions and velocities of the UAVs even when  OTFS channel estimation accuracy is limited because of the low resolution of the delay and Doppler shift grid.  
\aletext{Our results show a 10-fold improvement of the RMSE over the maximum delay discretization
error ($c/2B$) in the estimate of the length of the position vector, which translates to a 3- to 4-fold improvement on each 3D  component.
Taking advantage of the measurements from all the communication links between the UAVs, the remarkable advantage of our 
solution is thus the improvement of the $c/B$ resolution  of a multi-target radar system with a limited bandwidth $B$.}
As no other algorithm using delay-Doppler channel measurements is available in the literature, we demonstrated the excellent performance of TIP  against the Cram\'er Rao lower bound and a genie-aided version of TIP (with perfect knowledge of the maps).
Future work will extend our method to
include passive reflectors for terrestrial applications. 

{\color{black} In this work, we have assumed the pilot power to be sufficient to guarantee that all paths are resolved, i.e., that all lists $\Dc_{i,j}$ and $\Vc_{i,j}$ contain $N-2$ nLoS paths. The case where lists are incomplete,
 e.g., due to some paths being too weak, is a direction we are currently pursuing, although, determining the UAV’s identities in the incomplete lists is very challenging.}

\appendices
\section{Derivation of~\eqref{eq:v_ijk}\label{app:velocity}}
Let $\vec{P}_i$ be the position vector of UAV $i$ at time $t=0$ and  $\vec{V}_i$ its instantaneous velocity vector. Then, for small $t$, the position of $i$ can be written as
$\vec{P}_i(t) {=} \vec{P}_i + t \vec{V}_i + o(t^2)$
or, in terms of their components, $\pv_i(t) {=} \pv_i + t \vv_i + o(t^2)$.

The velocity of UAV $j$, as observed by $i$ through the reflexion on $k$, is the derivative w.r.t. time (evaluated at $t{=}0$) of the distance $s_{i,j,k}(t)$ between $j$ and $i$ through $k$. In turn, this distance is the sum of two contributions: the distance between $j$ and $k$, and the distance between $k$ and $i$. Hence, 
\[  
s_{i,j,k}(t) = |\pv_j(t)-\pv_k(t)| + |\pv_k(t)-\pv_i(t)|\,. 
\]
By using the chain rule, for any $n$ and $m$, we get
\begin{equation} 
\frac{\dd |\pv_n(t){-}\pv_m(t)|}{\dd t}\Bigg|_{t{=}0} 
= (\vv_n-\vv_m)\Tran \frac{\pv_n-\pv_m}{|\pv_n-\pv_m|}\,.
\end{equation}
It follows that the expression for $\omega_{i,j,k}$ is given by~\eqref{eq:v_ijk}.

\end{document}